\documentclass[12pt,twoside,fleqn]{iopart}
\usepackage{graphicx}

\def\gA{g_{\mbox{\tiny A}}}

\def\gV{g_{\mbox{\tiny V}}}

\def\sss{\scriptscriptstyle}

\def\endauthors{}
\def\authors#1\endauthors{#1}

\def\be{\begin{equation}}
\def\ee{\end{equation}}
\def\br{\begin{eqnarray}}
\def\er{\end{eqnarray}}
\def\brn{\begin{eqnarray*}}
\def\ern{\end{eqnarray*}}
\def\rf#1{{(\ref{#1})}}

\def\etal{{\it et al., }}

\def\ie{{\em i.e., }}

\def\lbar{\mbox{$\lambda$\kern-0,450em \vrule width0,35em height1,252ex
depth-1,21ex \kern0,051em}}
\def\dbar{\mbox{d\kern-0,347em \vrule width0,3em height1,252ex depth-1,21ex
\kern0,051em}}
\def\Dbar{\mbox{D\kern-0,735em \vrule width0,3em height0,86ex depth-0,81ex
\kern0,40em}}

\def\b {{\beta}}
\def\e {{\epsilon}}

\def\s {{\sigma}}

\def\n {{\nu}}

\def\ba#1{\begin{array}{#1}}
\def\ea{\end{array}}

\def\be{\begin{equation}}
\def\ee{\end{equation}}
\def\br{\begin{eqnarray}}
\def\er{\end{eqnarray}}
\def\brn{\begin{eqnarray*}}
\def\ern{\end{eqnarray*}}
\def\bit{\begin{itemize}}
\def\eit{\end{itemize}}
\def\bnu{\begin{enumerate}}
\def\enu{\end{enumerate}}
\def\x{\times}
\def\={{\simeq}}

\def\go{\rightarrow  }

\def\rf#1{{(\ref{#1})}}

\def\nn{\nonumber }

\def\2q{{{\{}2{\}}_q}}
\def\3q{{{\{}3{\}}_q}}

\def\sss{\scriptscriptstyle}

\begin{document}
\title{Gross Theory Model for Neutrino-Nucleus Cross Section}

\author{A. R. Samana$^{1,2,\dag}$, C. A. Barbero$^{3,4,*}$, S. B. Duarte$^2$, \\
        A. J. Dimarco$^5$ and  F. Krmpoti\'c$^{6,7}$}
\address{$^1$Department of Physics Texas A\&M University-Commerce, P.O.Box 3011 ,
Commerce, Texas, USA}
\address{$^2$Centro Brasileiro de Pesquisas F\'{\i}sicas,
        Rua Dr. Xavier Sigaud 150, CEP 22290-180, Rio de Janeiro-RJ, Brazil}
\address{$^3$Instituto de F\'isica La Plata, CONICET,
        1900 La Plata, Argentina}
\address{$^4$Departamento de
        F\'isica, Facultad de Ciencias Exactas, Universidad Nacional de La Plata, C.C. 67, 1900 La Plata, Argentina}
\address{$^5$Departamento de Ci\^encias Exactas e Tecnol\'ogicas,
        Universidade Estadual de Santa Cruz,
        CEP 45662-000 Ilhe\'us, Bahia-BA, Brazil}
\address{$^6$Departamento de F\'isica Matem\'atica, Instituto
        de F\'isica da Universidade de S\~ao Paulo, Caixa Postal 66318,
        05315-970 S\~ao Paulo-SP, Brazil}
\address{$^7$Facultad de Ciencias Astron\'omicas y
        Geof\'isicas, Universidad Nacional de La Plata, 1900 La Plata,
        Argentina.}
\ead{$^{\dag}$ arturo@cbpf.br\\\hspace{1.08cm} $^{*}$ barbero@fisica.unlp.edu.ar}

\begin{abstract}
The nuclear gross theory, originally formulated by Takahashi and
Yamada for the $\beta$-decays, is applied for the
electronic-neutrino nucleus reactions, employing a more realistic
description to the energetics of the Gamow-Teller resonances. The
model parameters are gauged from the most recent experimental
data, both for $\beta^-$ decay and electron-capture,
separately for even-even, even-odd, odd-odd, odd-even nuclei.
The numerical estimates for neutrino-nucleus cross sections agree
fairly well with previous evaluations done within the framework of
microscopic models. The formalism presented here can be extended
to the heavy nuclei mass region, where weak processes are
quite relevant, which is of astrophysical interest because of its
applications in supernova explosive nucleosynthesis.

\end{abstract}

\pacs{21.60.-n, 21.10.-k}
\maketitle

\section{Introduction}

The nucleosynthesis of heavy elements is only understood if stellar
reactions take place in regions of nuclear chart far away from the
$\beta$-stability line, involving a large number of unstable or even
exotic nuclear species for which  the  experimental data are very
scarce. For instance, the steps of nucleosynthesis  in the
$r$-process occurs out to and just along the neutron drip line where
many of principal nuclear properties are still unknown.
Great  theoretical and experimental efforts have been invested in
the last decades in order to describe the nuclear properties of
different species  along the $\beta$-stability line, as well as
those of  exotic nuclei involved in explosive nucleosynthesis
processes \cite{Gor05,Hil02,Bor03}.

The theoretical models can be separated generically into: i) the
macroscopic models which describe the global nuclear properties
\cite{Tak69,Koy70,Kon85,Tac90,Tac92,Nak97,Qia97}, and where
special attention is paid to the {\it gross theory of the
$\beta$-decay} (GTBD); and ii) the microscopic formalisms  \ie the
shell model or RPA based calculations
\cite{Qia97,Lan99,Bor00,Gor02} where detailed nuclear structure of
each species is considered.

The GTBD was first proposed by Takahashi and Yamada ~\cite{Tak69} nearly
forty years ago to describe the global properties of allowed
$\beta$-decay  processes. It is essentially a parametric model,
which attempted to combine the single-particle and statistical
arguments in a phenomenological way.
Afterwards,  different versions of the 'gross theory' have been developed
and used for practical applications
very frequently \cite{Koy70,Kon85,Tac90,Tac92,Nak97}.
This is due to: i) their simplicity when compared with the hard
computational work involved in the implementation of the microscopic
models, and ii) their capability to reproduce the available
experimental data, and to be extrapolated later on to unknown nuclei
far away from the $\beta$-stability line. In fact, as these
theoretical approaches account systematically and fairly well for
the properties of stable nuclei, they have been extensively applied
to describe:  1) the $\beta$-decay half-lives and other nuclear
observables participating in the $r$-process, and 2) the properties
of a great number of  exotic nuclei that are involved in the
nucleosynthesis.

It also should be mentioned that the gross-theory approach has been also used by
N. Itoh \etal in Refs.~\cite{Ito77}
 for the calculation of the total capture of a neutrino
 by $^{37}$Cl, $^{16}$O, $^{20}$Ne
and $^{56}$Fe nuclei, which are used in the detection of solar neutrinos.

The aim of the present work is twofold. First, motivated by the
simplicity of the original GTBD, we use it  to  evaluate the
half-lives of  allowed weak-transitions ($\beta$-decay  and
electron-capture) in nuclei with $A < 70$, which are of major
importance for  the presupernova collapse processes. We also analyze
the consequences of employing  a more realistic estimate for the
energetic of the Gamow-Teller resonance (GTR) than in the previous
works. This will lead us to a new trend for the adjustable parameter
related to the energy spread of the GTR caused by
the spin dependent part of the nuclear force. Second, we use the same
gross-theory approach to describe the nuclear neutrino capture over
a great number of nuclei involved in presupernova
structure with the purpose to extend in the future the calculation
to $r$-process in neutrino rich environment.
Since  within the stellar conditions no  experimental data exist,
our  results are confronted with those achieved in the
framework of microscopic approaches.

The paper is organized as follows. In Sect. 2 we  briefly sketch
the conventional gross-theory for  nuclear $\beta$-decay and
electron capture rate. In Sect. 3 we introduce the gross theory  for the evaluation of
the neutrino-nucleus reaction cross-section. The single-particle
strength functions are discussed in Sects. 4 and 5, together with
the estimate of the GTR energy and the procedure used to
derive the corresponding spread of the transition strength.
In Sect. 6 we exhibit and discuss our results.
Summarizing conclusions and future extension of the present work are drawn in
Sect. 7.

\section{Gross theory of nuclear beta decay (GTBD)}

The GTBD permits to evaluate the half-lives of $\beta^\pm$-decay
and the rates for electron capture weak processes. As an example,
we briefly sketch here the original GTBD~\cite{Tak69} for the
decay $(Z,A)\go (Z+1,A)+e^-+\tilde{\nu}$. The total  rate for
allowed transitions is written (in natural units) as
\br
\lambda_\b=\frac{G_{\sss F}^2}{2\pi^3}\int_{-Q_\b}^0dE\left[
g_{\sss V}^2\left|{\cal M}_{\sss F}(E)\right|^2+g_{\sss A}^2
\left|{\cal M}_{\sss GT}(E)\right|^2\right]f(-E), \label{2.2}\er
where  $G=(3.034545\pm 0.00006)\x 10^{-12}$ is the Fermi weak
coupling constant, $g_{\sss V}=1$ and $g_{\sss A}=-1$ are,
respectively, the vector and axial-vector coupling constants
 \footnote{Finite nuclear
size effects are incorporated via the dipole form factor
$g\rightarrow g \left( \frac{\Lambda^2}{\Lambda^2+k^2} \right)$
where $k$ is the momentum transfer and $\Lambda=850$ MeV the
cutoff energy. }.
The argument of the matrix element ($E$) is the
transition energy measured from the  parent ground state.
Note that  the true $\b$-decay
transition energy is $E_\b=E_e+E_\nu=-E>0$.
The usual integrated dimensionless Fermi
function \cite{Fee50,Ros60}, $f(E)$,  is evaluated from
the approximated formulas given in Ref.\cite{Tak69} that are correct
up to $\sim 10\%$ for standard decays.
The  $Q_\b$-value is the difference between  neutral atomic masses
of parent and daughter nuclei:
\br
Q_{\b^-}=M(A,Z)-M(A,Z+1)=B(A,Z+1)-B(A,Z)+m(nH)
\label{2}\er
with $B(A,Z)$ and  $B(A,Z+1)$ being the
corresponding nuclear binding energies, and $m(nH)=m_n-m(^1 H)=m_n-m_p-m_e=0.782$ MeV.
The masses were obtained in the
same way as in Ref. \cite{Tac90}. This means that, when  available,
they are taken   from  Wapstra-Audi-Hoekstra  mass table \cite{Wap88}
and, otherwise, they are determined from  Tachibana-Uno-Yamada
semi-empirical mass formula \cite{Tac88}.

The squares of the Fermi (F) and Gamow-Teller (GT)
matrix elements are determined as:
\be
\left|{\cal M}_{\sss
X}(E)\right|^2 =\int_{\e_{min}}^{\e_{max}} D_{\sss X}(E,\e)
W(E,\e) \frac{dn_1}{d\e} d\e,~~~~~~  {\mbox {for $X= F, GT$}}.
\label{3}\ee
Here, $\e_{min}$ is the lowest  single-particle energy
of the parent nucleus and   $\e_{max}$
is the energy of the highest occupied state.
The one-particle level density (proton or neutron), $dn_1/d\e$,
is determined by Fermi gas model for the parent nucleus,
and the weight function $W(E,\e)$, constrained
by $0\leq W(E,\e) \leq 1 $, takes into
account the Pauli blocking. Finally,  $D_{\sss X}(E,\e)$,
normalized as
$\int^{\sss +\infty}_{\sss {-\infty}} D_{\sss X}(E,\e) dE=1$,
is the probability that a nucleon with single-particle energy $\e$ makes
a $\beta$-transition. As in Ref.~\cite{Tak69} we neglect the
$\e$-dependence, \ie it is assumed that all nucleons have the same
decay probability, independently of their energies $\e$,  $D_{\sss
X}(E,\e)\equiv D_{\sss X}(E)$.
The GTBD characterizes this $D_{\sss X}(E)$ through their
energy weight moments (for example, in \cite{Ito77} these
expressions were written explicitly).

The dependence on the odd-even proton and neutron numbers in the
daughter nucleus is introduced through the values for the  pairing
gap $\Delta$ and the  single-particle level spacing $d$. In the
present work we adopted those from Ref. \cite{Koy70}. More
details on the probability function $D_{\sss X}(E)$ are given in
Sect. 4.

The original GTBD \cite{Tak69} has been gradually improved
\cite{Kon85,Tac90}, and nowadays  we have two new versions:
the first is named the 2nd generation gross theory (GT2), and
the second is the so called semi-gross theory (SGT) in which some
parts of nuclear shell effects are considered. The most recent GT2
and SGT approaches use an updated mass formula, and they better account
for the shell and pairing effects \cite{Tac90,Nak97}.

\section{Gross theory of nuclear neutrino capture (GTNC)}

In the most recent versions of $r$-processes nucleoshynthesis
in supernova, one considers that these processes take place
on the surface of a protoneutron star during the supernova collapse.
The nuclei are exposed there to a thermal flux
$\Phi_\nu(E_\nu)$ of $\nu_e$
with energy $E_\nu$, which causes the reaction
$\nu_e+(Z,A)\go(Z+1,A)+e^-$, with
cross-section~\cite{Bor00,Qia97,Krm05}
\br
\langle \sigma_\nu
\rangle= \int_{E_{th}}^{\infty} \Phi_\nu(E_\nu) \sigma_\nu(E_\nu)
dE_\nu,
\label{4}\er
where $E_{th}$ is the reaction energy
threshold, which is equal to the $Q_\b$-value for stable nuclei and
zero for unstable cases.
For $\Phi_\nu(E_\nu)$ we take a zero-chemical potential
Fermi-Dirac distribution
\be
\Phi_\nu(E_\nu)= \frac{{\cal
N}}{T_\nu^3}\frac{E_\nu^2}{e^{E_\nu/T_\nu}+1},
\label{5}\ee
where $T_\nu$ is the neutrino temperature,  and ${\cal N}$ is
the normalization constant of the spectrum \cite{Bor00}.

The evaluation of the  $\nu_e$-nucleus cross-section
$\sigma_\nu(E_\nu)$, in a
neutrino-rich environment, must be
consistent with the procedure employed in calculating
the $\beta$-decay rates.
The allowed transition
approximation (see \cite[Eq.(2.19)]{Krm05})
\br
\s_\nu (E_\nu)&=&
\frac{G^2}{\pi} \int_0^{E_\nu-m_e} p_e E_e F(Z+1,E_e) \left[
\gV^2\left|{\cal M}_{\sss F}(E)\right|^2 + \gA^2\left|{\cal
M}_{\sss GT}(E)\right|^2 \right] dE,
\nn \\
\label{6}\er
can be applied  for relatively small momentum transfer.  The integration
covers all possible nuclear
states allowed by the selection rules, and the integration limits
are determined from the energy conservation condition.
When the energies
are measured from the ground state of the  parent nucleus $(Z,A)$,
this condition reads
\br
E_\nu+M(Z,A)=E_e+M(Z+1,A)+Q_{\b^-}+E,
\label{7}\er
where $E=E_\nu-E_e>0$ is the excitation energy of daughter nucleus
$(Z+1,A)$, and $F(Z,E)$ is the usual scattering Fermi function
which takes into account the Coulomb interaction between the
electron and the nucleus.

\section{Single-particle strength functions}

A key element in the gross theory is the
single-particle strength probability function $D_{\sss X}(E)$. The
successive improvements of the theory have used gaussian-, exponential-, and
lorentzian-type functions \cite{Tak69,Tac90}.
The sec-hyperbolic functions have been employed in the GT2 \cite{Tac90}.
Here we will mainly adopt the gaussian-like behavior for
the transition strengths.
To illustrate that the calculations are rather
independent of the functional form adopted for $D_{\sss X}(E)$,
a comparison will be done between the results obtained with the
gaussian-like distribution
\be
D_{\sss X}(E)= \frac{1}{\sqrt{2\pi
\sigma_{\sss X}}}e^{-(E-E_{\sss X})^2/(2 \sigma_{\sss X}^2)},
\label{8}\ee
and those calculated with the lorentzian-type
strength function
\be
D_{\sss X}(E)=\frac{\Gamma_{\sss X}}{2\pi}
\frac{1}{(E-E_{\sss X})^2+(\Gamma_{\sss X}/2)^2} ~ \cdot
\label{9}\ee
Here $E_{\sss X}$ is the resonance energy, $\s_{\sss X}$ is
the standard deviation, and the other quantities are defined
as in Ref.~\cite{Tak69}.

When isospin is a good quantum number the total Fermi strength
$\int \left|{\cal M}_{\sss F}(E)\right|dE=N-Z$ is carried entirely
by the isobaric analog state (IAS) in the daughter nucleus. However,
because of the Coulomb force, the isospin is not a good quantum
number and this  leads to the energy splitting of the
Fermi resonance. We  will use the  estimates introduced by
Takahashi and Yamada \cite{Tak69}, namely
\br
E_F&=&\pm(1.44 ZA^{-1/3}-0.7825)~{\mbox{MeV}}; ~~~~~{\mbox{for $\beta^\pm$
decay}},
\nn\\
\s_F&=&0.157 Z A^{-1/3}~{\mbox{MeV}}.
\label{10}\er
The total GT strength in the $(\n_e,e^-)$ channel is given by the
Ikeda sum rule $\int \left|{\cal M}_{\sss GT}(E)\right|dE\cong 3(N-Z)$,
but its distribution  cannot be established by general
arguments, and therefore must be either calculated or measured.
Charge-exchange reactions $(p,n)$ have demonstrated that most of
the strength is accumulated in a broad resonance near the
IAS~\cite{Hor80}. In fact, even before these measurements
have been performed, Takahashi and Yamada \cite{Tak69} have used
the approximation
\be
E_{GT}\cong E_F,
\label{11}\ee
while $\sigma_{GT}$ is expressed as
\be
\sigma_{GT}=\sqrt{\sigma_F^2+\sigma_N^2},
\label{12}\ee
with $\sigma_N$ being the energy spread caused by the spin
dependent nuclear forces.

For the Fermi transitions we  use the relation \rf{10}. Yet,  for
the GT resonance, instead of employing the approximation \rf{11}, we
use the estimate
\be
E_{GT}=E_F+\delta; \hspace{.6cm}
\delta=26A^{-1/3}-18.5(N - Z)/A~~~~ \mbox{MeV},
\label{13}\ee
obtained by Nakayama {\it et al.} \cite{Krm82} from the analytic
fit of the $(p,n)$ reaction data of nuclei near stability line
\cite{Hor80}, where $\delta$ is positive. For the standard deviation
$\sigma_{GT}$ we  preserve the expression \rf{12}, and $\sigma_N$ is
treated as an adjustable parameter.
Note that the two terms of $\delta$ in \rf{13} have well defined physical
interpretations. The first one is due to the $SU(4)$ symmetry
breaking imposed by the spin-orbit coupling, and it is of the same
order of magnitude as the Bohr-Mottelson estimate for the spin-orbit
splitting ($\Delta_{ls}\cong 20 A^{-1/3}$ MeV), obtained from the
approximation $l\cong A^{1/3}$~\cite{Bor75}. The second term is
responsible for the partial restoration of the  $SU(4)$ symmetry,
having the same mass and charge dependence as the difference between
the energy shifts produced by the GT and Fermi residual
interactions.
We remark that the Eq.~\rf{13} is frequently used in the study of
$r$-process in neutron rich nuclei
\cite{Qia97,Ben02,Lan00,Sur98,Kar98,Sur95,Kar94,Dea94,Kar91,COO84,Hek00}.
There $\delta<0$, and therefore the GTR falls below the IAS,
as happens in the shell-model calculation~\cite{Qia97}.
\footnote{
Occasionally is used the fit \cite{Tac90}
\[
E_{GT}=E_F+\delta'; \hspace{.6cm}  \delta'=6.7-30(N - Z)/A
 \hspace{.6cm}~~~~
\mbox{MeV},\]
which also reproduces satisfactorily the stable nuclei.
The second term of $\delta'$ is  interpreted
in the same way as that of $\delta$ in \rf{13}, but the first term here
does not have any direct physical significance.}

\section{ Fitting Procedure}

Another important aspect in implementing the GTBD is
the choice of the $\chi^2$-minimization method
that is used  to derive the width parameter $\sigma_N$.
In the original work of Takahashi and Yamada \cite{Tak69} is
minimized the quantity
\be
\chi_A^2=\sum_{n=1}^{N_0}\left[\log\left(\tau_{1/2}^{cal}(n)
/\tau_{1/2}^{exp}(n)\right)\right]^2,
\label{14}\ee
where $N_0$ is the number of experimental $\b$-decay
half-lives,  $\tau_{1/2}^{exp}$, fulfilling the conditions:
1) the branching ratio of the allowed transitions exceeds  $\sim 50\%$ of
the total $\b$-decay branching ratio, and 2) the ground state
$Q$-value is $\geq 10 A^{-1/3}$ MeV.

In the present  work $\sigma_N$ is determined through the
minimization of the function
\be
\chi_B^2=\sum_{n=1}^{N_0}\left[\frac{\log(\tau_{1/2}^{cal}(n)/
\tau_{1/2}^{exp}(n))} {\Delta\log(\tau_{1/2}^{ exp}(n))}\right]^2,
\label{15}\ee
where
\be
\Delta\log (\tau_{1/2}^{exp}(n))\equiv
\left|\log[\tau_{1/2}^{exp}(n)+\delta\tau_{1/2}^{exp}(n)]
      -\log[\tau_{1/2}^{exp}(n)]\right|,
\label{16}\ee
and $\delta\tau_{1/2}^{exp}$ is the experimental error. Thus,
the $\chi^2_B$-function  reinforces the contributions
of  data with small experimental errors.
Moreover, we perform different fittings for
even-even, odd-odd, odd-even and even-odd nuclei.
Needless to say that for  $\tau_{1/2}^{exp}$ we use here
the most recent data \cite{Wallet}, instead of those that were
available when the GTBD has been
formulated~\cite{Tak69}. The condition $\log ft \le 6$
is imposed  to include only the allowed $\beta$-decays.

\section{Numerical results and discussion}

\subsection{$\beta^-$ decay and electron-capture half-lives}

For the single-particle
strength probability function $D_X(E)$ we adopt the gaussian-like
behavior \rf{8} in most of the calculations.
The corresponding values of the adjustable parameters  at the minimal
value of the $\chi^2$-function, $\chi^2_{min}$, are listed in
Table I for the four different parity families of nuclei. They are
labeled as $\sigma_N^*$ and $\sigma_N$, when  for $E_{GT}$
are used, respectively, the Eqs.~\rf{11}
and \rf{13}. One sees that $\sigma_N$ is always larger than $\sigma_N^*$,
which means that the effect of using  more realistic energies $E_{GT}$ is reflected
in the increase of the standard deviations.
The values of $\sigma_N^*$  derived in Ref. \cite{Tak69} are
exhibited parenthetically in Table I. It is important
to point out that the difference between the old and new values for
$\sigma_N^*$ does not comes from the fitting procedure itself, but
from the different samples of nuclei employed here for each
parity family.

\begin{figure}[t]
\vspace{-0.cm}
\centering
\begin{tabular}{cc}
{\includegraphics[width=8cm,height=10cm]{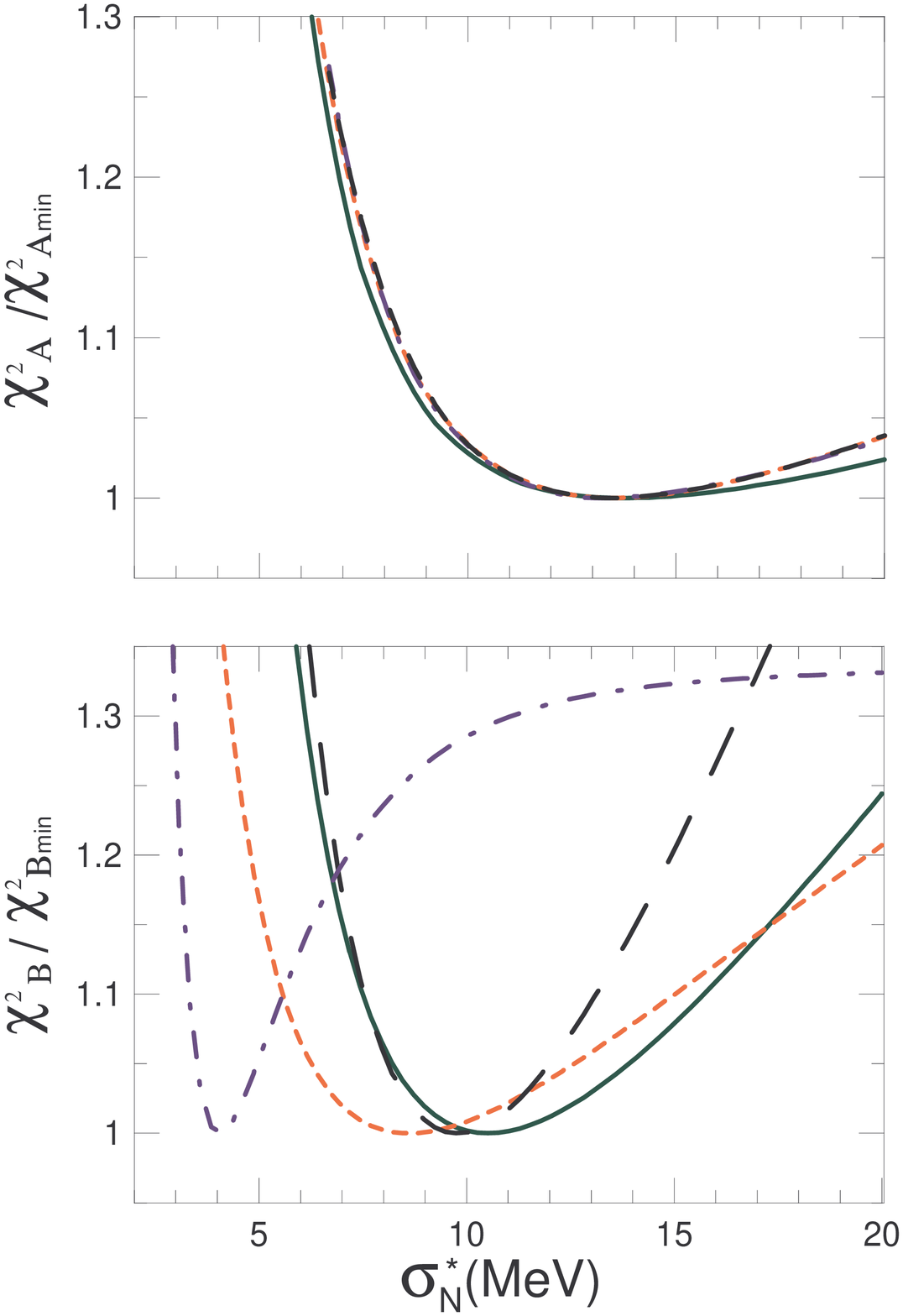}}
&\hspace{-1.cm}
{\includegraphics[width=8cm,height=10cm]{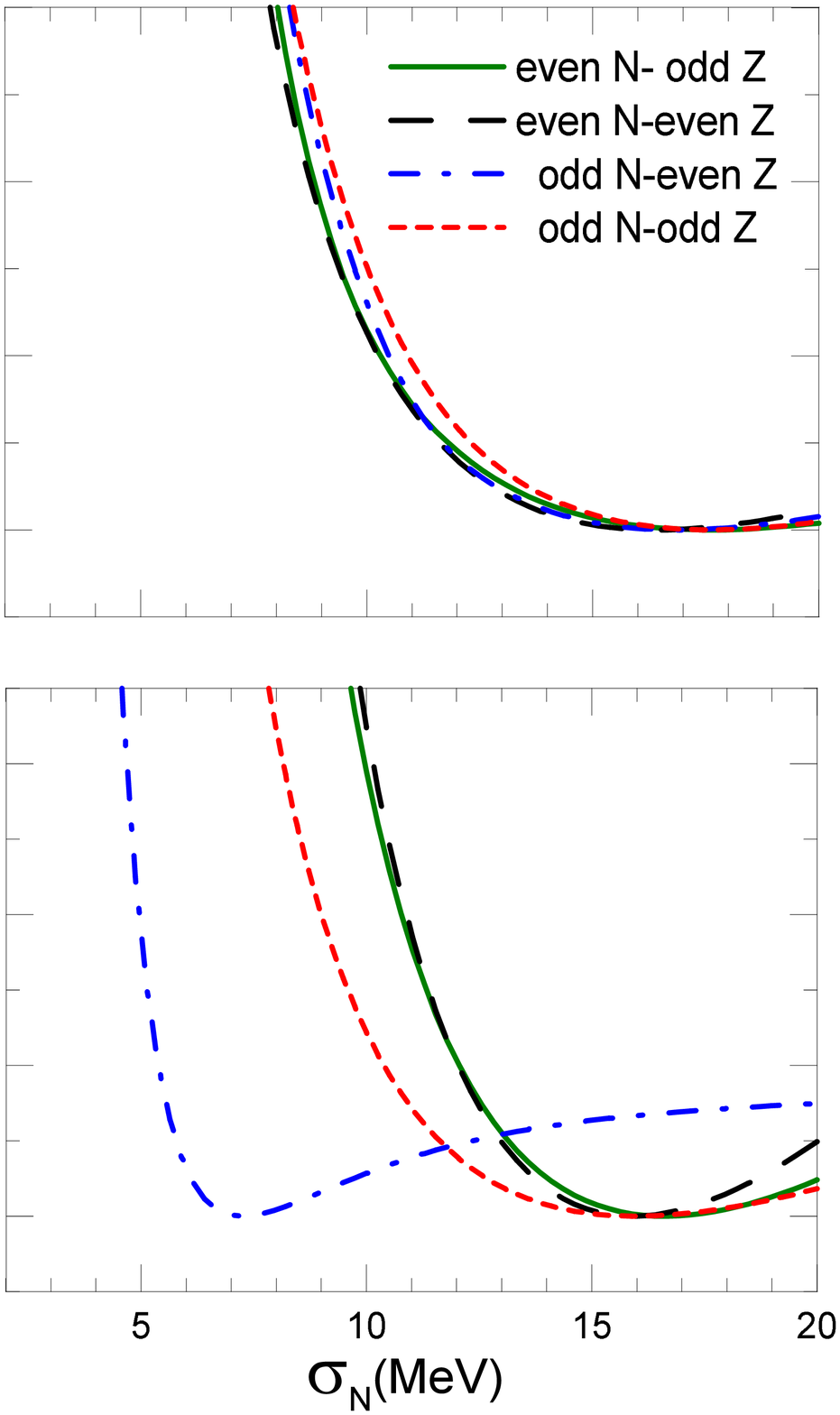}}
\end{tabular}
\caption{\label{figure1} Comparison between $\chi^2$ functions
(normalized to the minimum) for the $\beta^-$decay. Two types of
approximations were used for the  energy of the GTR: the left
panel shows the results obtained with the original estimate
\protect{\rf{11}}; the right panel includes the energy difference
between the GTR  and the IAS through the Eq. \rf{13}.}
\end{figure}

Figure \ref{figure1} shows the dependence of
$\chi^2/\chi^2_{min}$ on both: i) the energy of the
GTR (left panels for \rf{11}, and
right panels for \rf{13}), and  ii) the type of the minimization
function (upper panels for \rf{14}, and lower panels for \rf{15}).
We note that  the  $\chi^2_B$-functions  present
rather  pronounced minima when compared with those of
the $ \chi^2_A$-functions.
More, in most of the cases the $\chi^2_B$ minima are located
at smaller values of the standard deviations
than the  $\chi^2_A$ ones. This is  a direct  consequence
of including the experimental errors
in the minimization procedure of the $\chi^2_B$-function.

In order to estimate the average deviation of our results, we have
computed the mismatch factor $\eta$ defined as \cite{Tak69}
\be
\eta=10^{\sqrt{\chi^2/N_0}} ~~,
\label{17}\ee
showing their values for each $\sigma_N$ in
Table \ref{table:1}, and similarly the values
of $\eta^*$ corresponding to each $\sigma_N^*$. It can be observed
that the $\chi^2_B$ minimization procedure
considerably reduces the mismatch factor, in particular for odd-odd
family of nuclei.
Thus, we can say that the use of $\chi^2_B$-function modifies
$\sigma_N$ and leads to a better statistical agreement
between the theoretical results and the experimental data.

\begin{figure}[t]
\vspace{-1.cm}
\begin{center}
{\includegraphics[width=14cm,height=14cm]{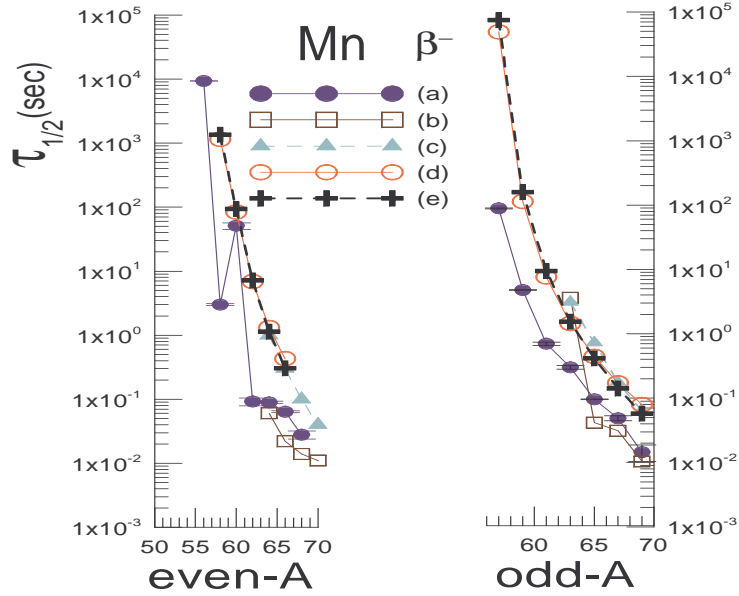}}
\end{center}
\vspace{-1.5cm}
\caption{\label{figure2}(Color online) Comparison
of $\beta^-$-decay half-lives for Mn: (a)
experimental; (b) ETFSI+CQRPA~ \cite{Bor00}; (c) ETFSI+GT2~
\cite{Tac90}; (d) GTBD with $E_{GT}$ from Eq. \rf{13}; and (e)
GTBD with $E_{GT}$ from Eq. \rf{11}. In both GTBD calculations
the gaussian type functions for $D_{F,GT}(E)$ was used.}
\end{figure}

Figure~\ref{figure2} compares the experimental
$\beta^-$-decay half-lives within  the Mn isotopic chain with
our results obtained for the $\sigma_N$
values from Table \ref{table:1}.
One can see that the GTBD overestimates the data.
However, it should be pointed out that this disagreement
is not characteristic of the GTBD, since other microscopic and
global models lead to similar results.
For instance, this is the case of:
a) the extended Thomas-Fermi plus Strutinsky integral
method combined  with the continuum
quasiparticle random phase approximation (ETFSI+CQRPA)~\cite{Bor00},
and b) the extended Thomas Fermi method combined with the semi-gross
theory (ETFSI+GT2)~\cite{Tac90}.

\begin{figure}[t]
\centering
\begin{tabular}{cc}
\vspace{-1.5cm}
{\includegraphics[width=7cm,height=9cm]{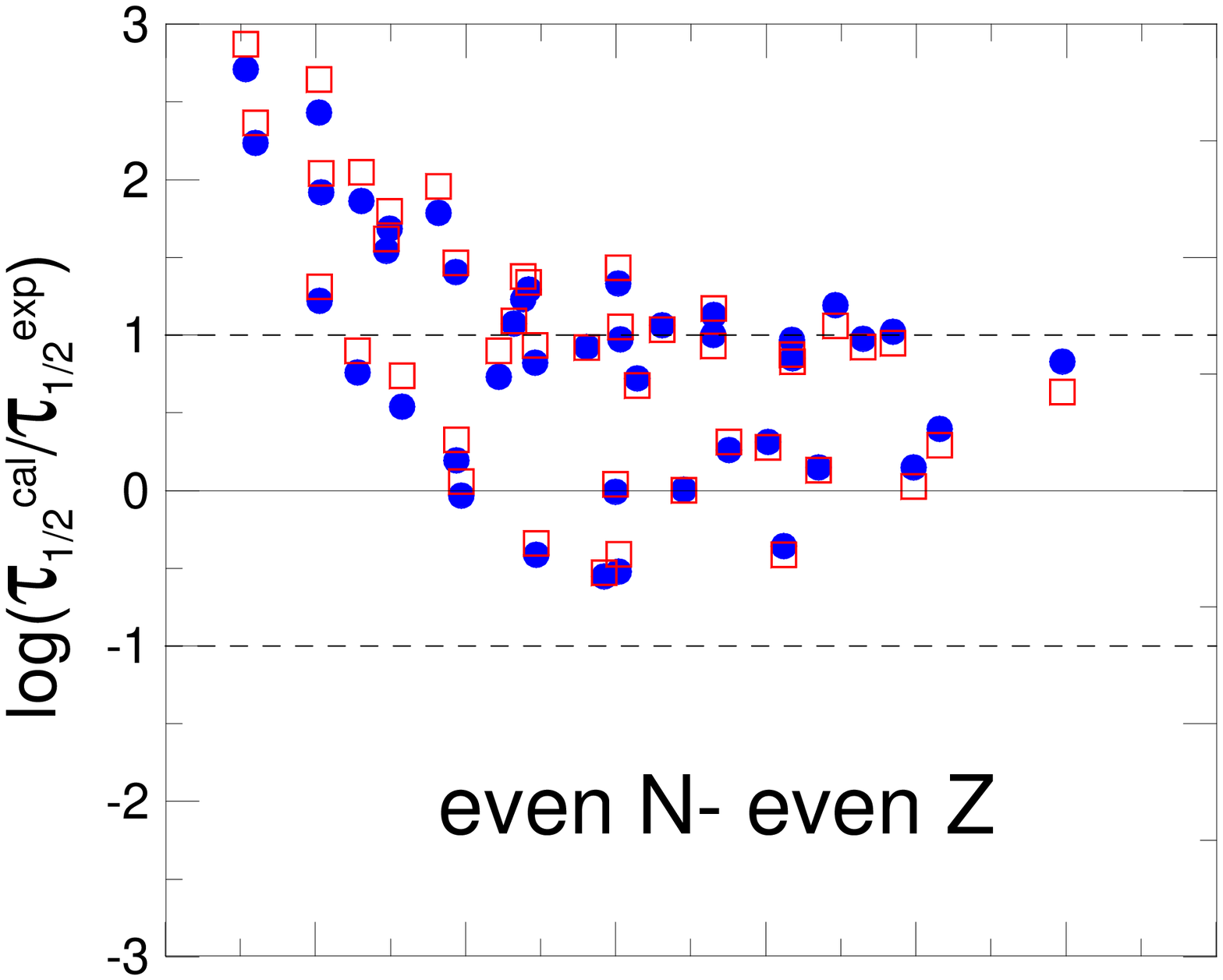}}
\vspace{-.25cm}
&
\vspace{-1.5cm}
{\includegraphics[width=7cm,height=9cm]{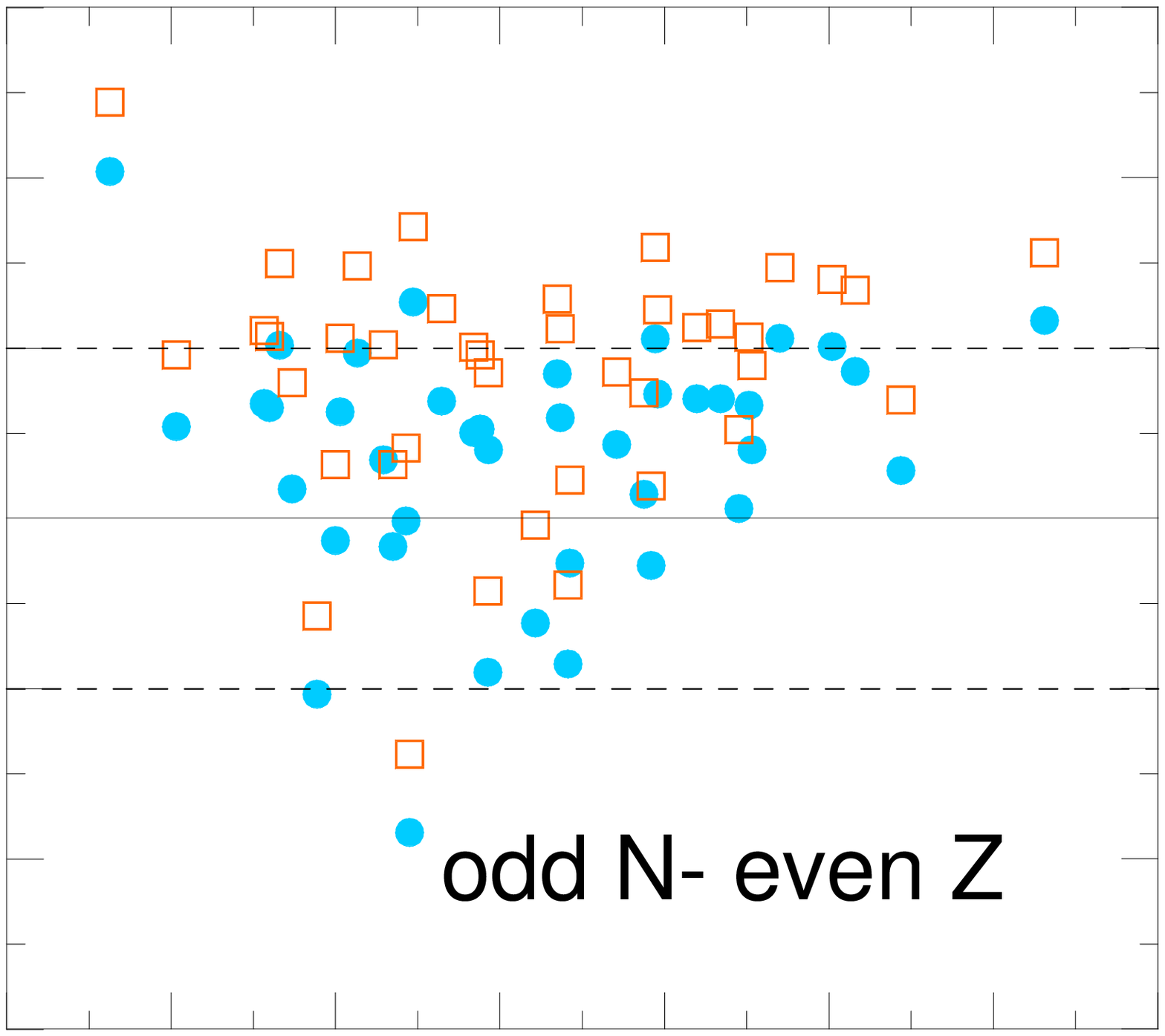}}
\vspace{-.25cm}
\\
{\includegraphics[width=7cm,height=9cm]{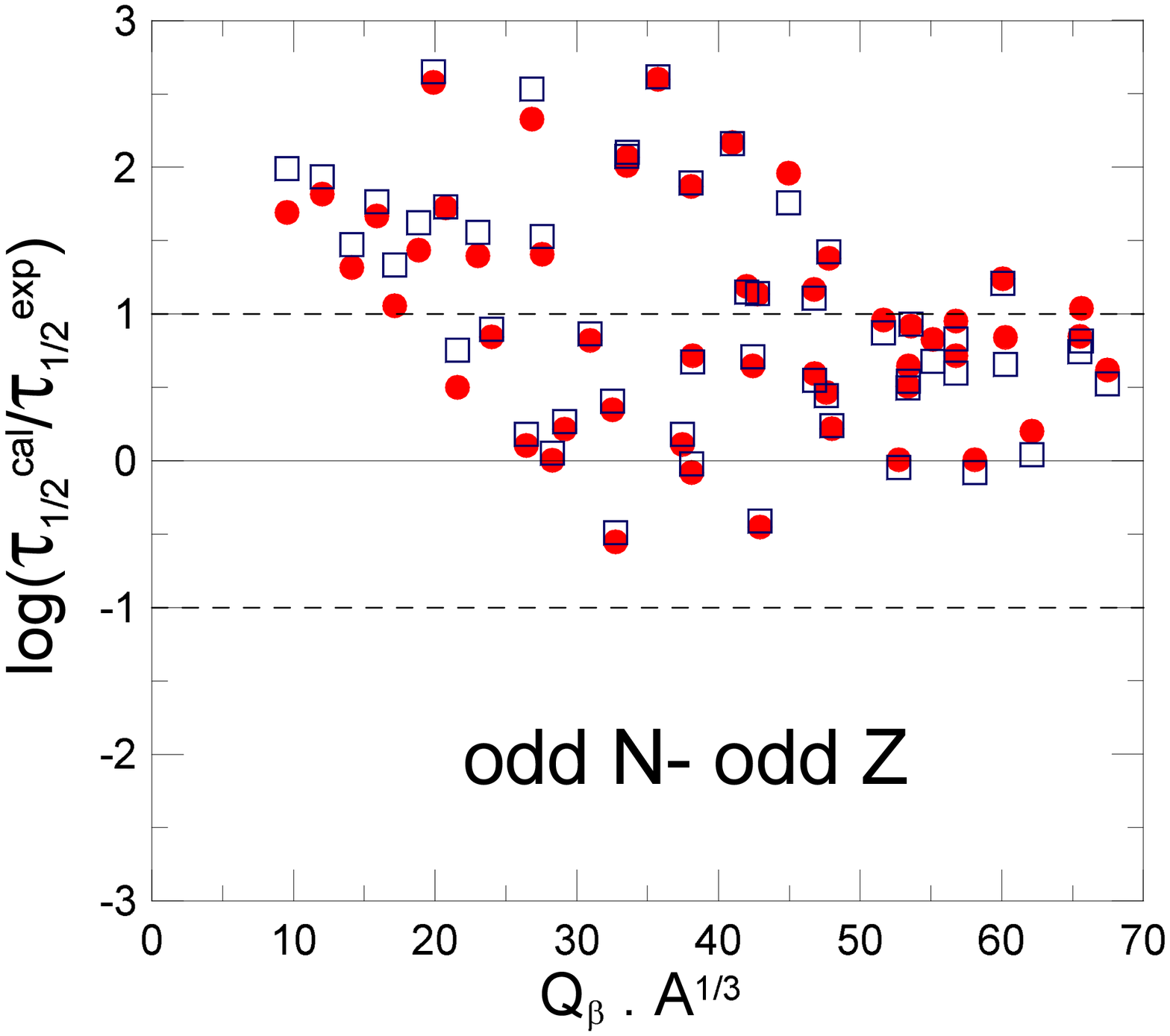}}
\vspace{-1.cm} 
&
{\includegraphics[width=7cm,height=9cm]{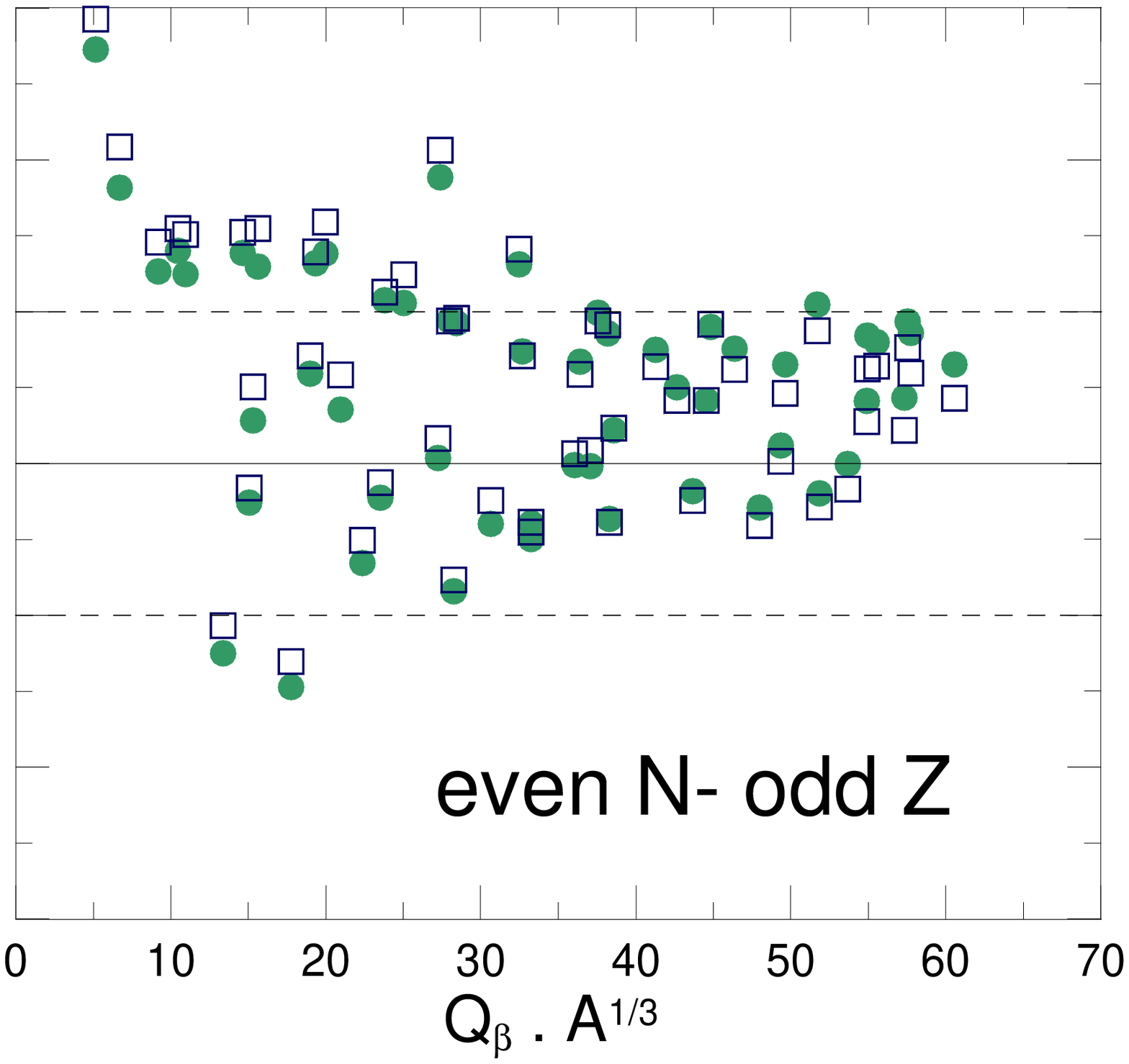}}
\vspace{-1.cm} 
\end{tabular}
\caption{\label{figure3}(Color online)
$log(\tau_{1/2}^{calc}/\tau_{1/2}^{exp})$ as a function of $Q_\beta
A^{1/3}$ for  $\beta^-$-decay of nuclei with $A<70$. Gaussian
functions were used for $D_{X}(E)$. We present the values obtained
with the approximations \rf{13} (filled circles) and \rf{11} (hole
squares) for $E_{GT}$.}
\end{figure}

Figure~\ref{figure3} shows the distribution of
$\log(\tau_{1/2}^{cal}/\tau_{1/2}^{exp})$, as a function of
$Q_\b A^{-1/3}$, for $\beta^-$-decay. We observe that the results  obtained with
Eqs.~\rf{11} and \rf{13} are
quite similar to each other for  same  parity families,
the first one being somewhat larger.
We can also see that for the odd-odd family a very good agreement
between theoretical and experimental results is obtained
for $Q_\beta A^{1/3} \geq 45$ MeV, while for the
other three  families
this happens already for $Q_\beta A^{1/3} \geq 40$ MeV.
Thus, as frequently mentioned in the literature
\cite{Tak69,Tac90,Nak97},  the best GTBD results are obtained for
heavy nuclei.

\begin{figure}[t]
\centering \vspace{-1.cm}
\begin{tabular}{cc}
{\includegraphics[width=8cm,height=9.cm]{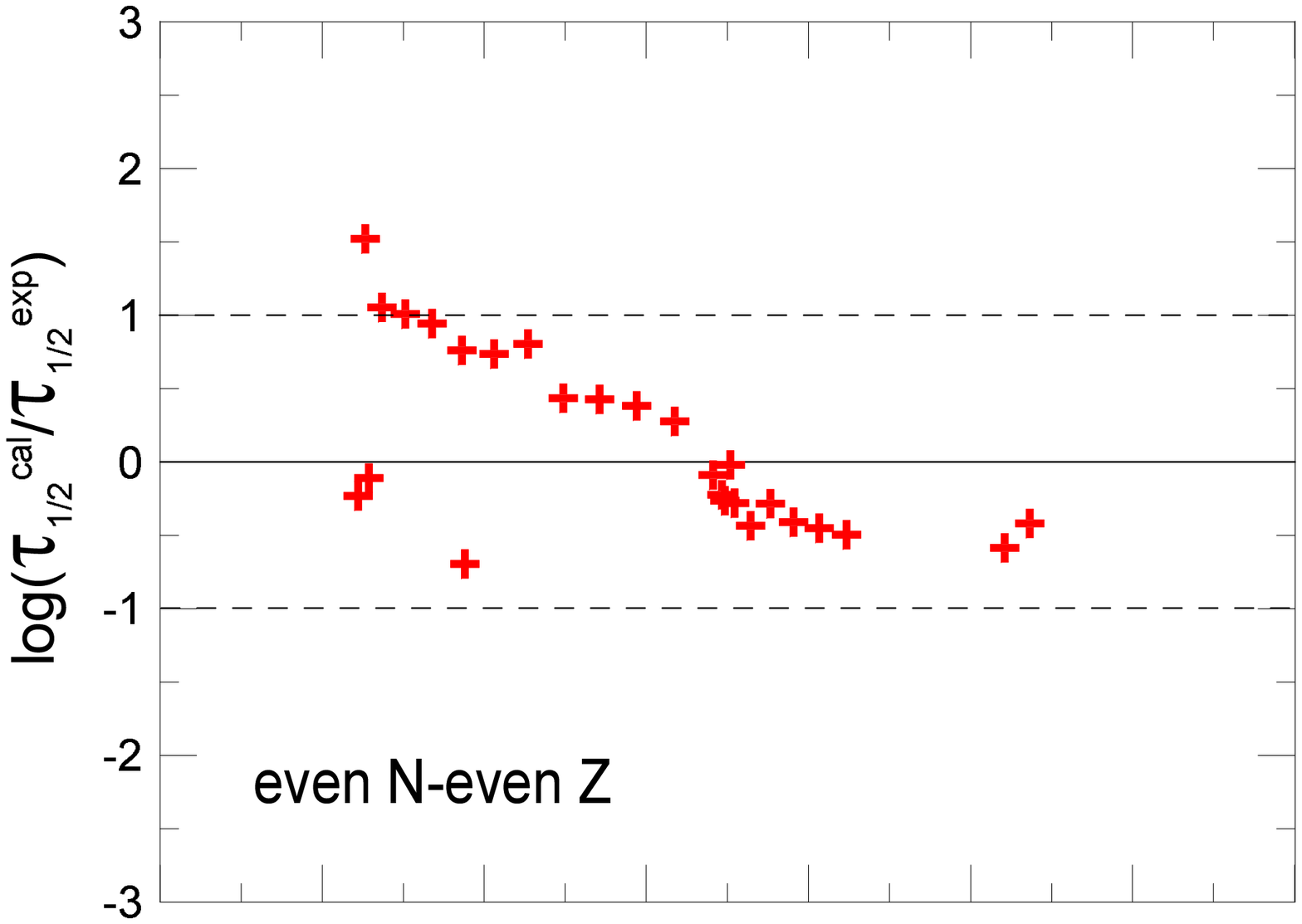}}& \hspace{-.8cm}
{\includegraphics[width=8cm,height=9.cm]{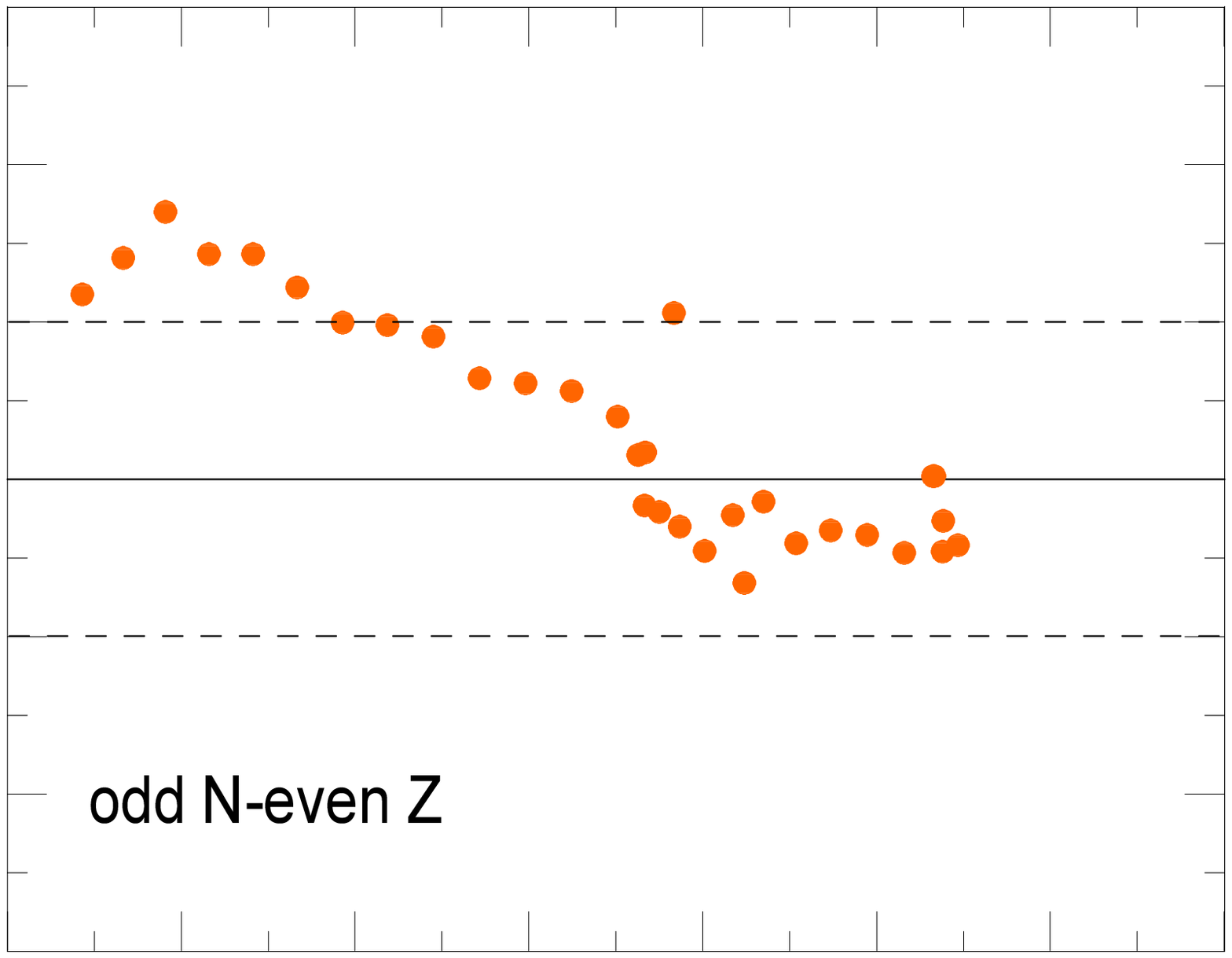}} \vspace{-4.5cm}
\\
{\includegraphics[width=8cm,height=9.cm]{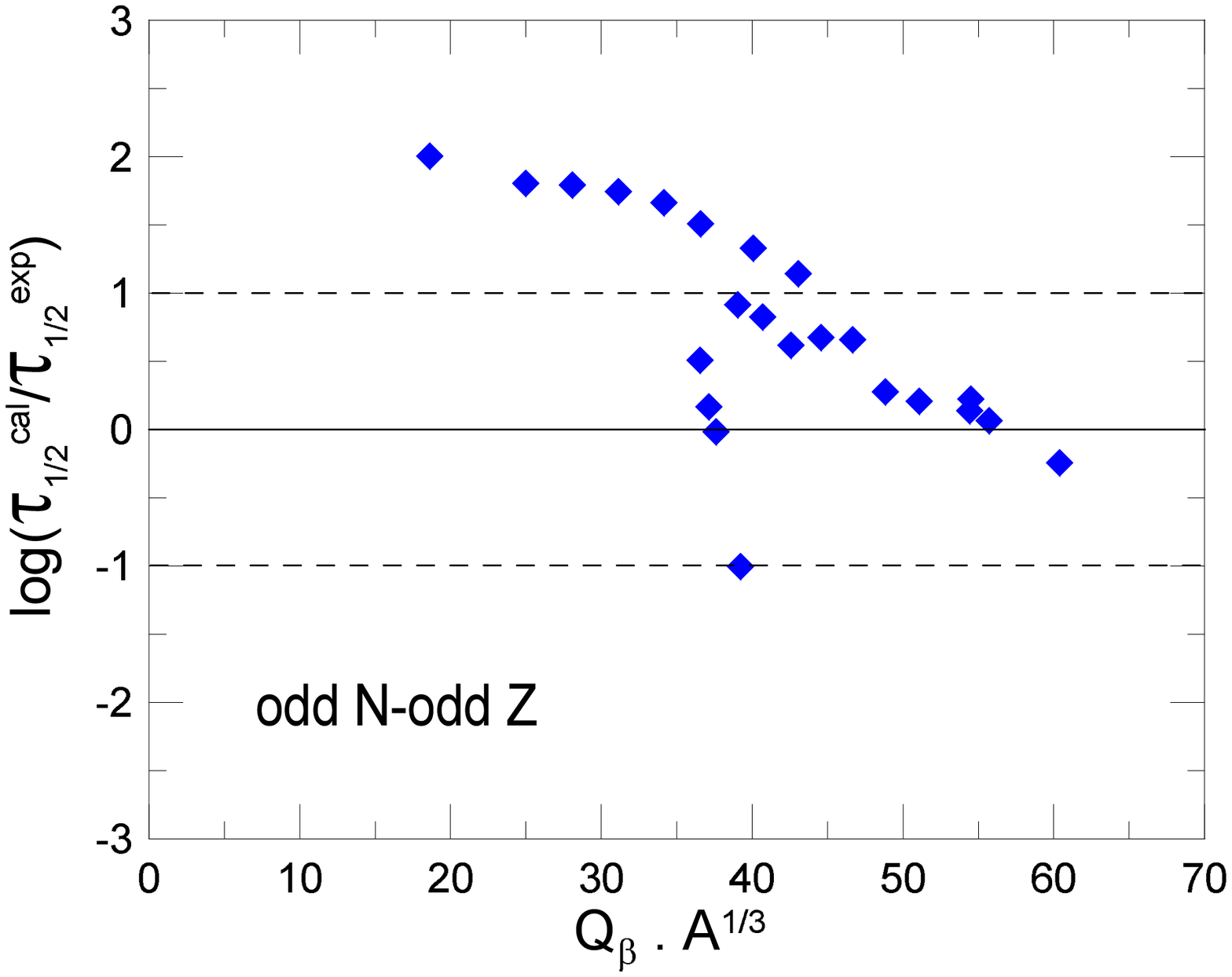}}& \hspace{-.8cm}
{\includegraphics[width=8cm,height=9.cm]{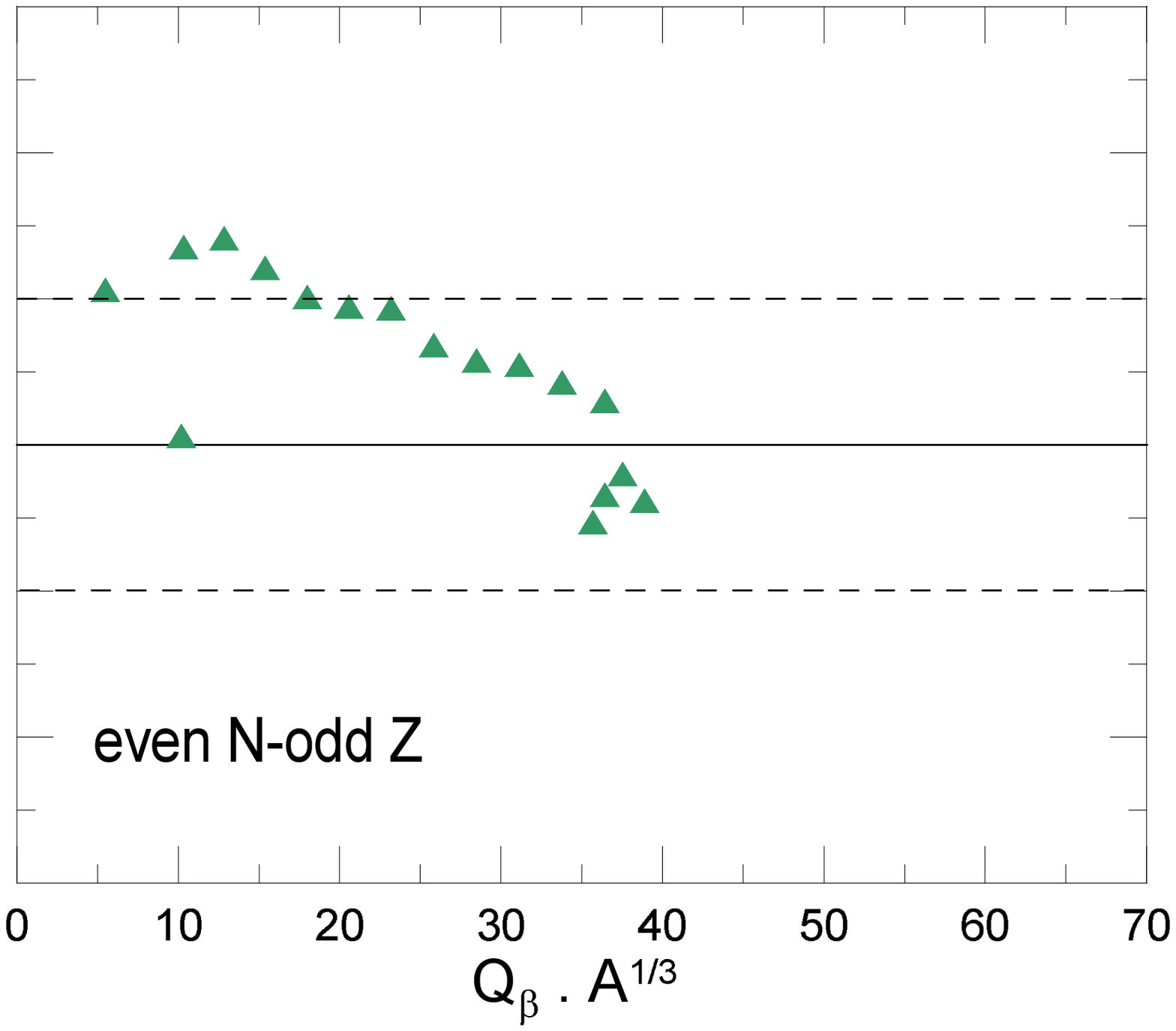}}
\end{tabular}
\vspace{-2.cm} \caption{\label{figure4} (Color online)
$log(\tau_{1/2}^{calc}/\tau_{1/2}^{exp})$ as a function of
$QA^{1/3}$ for electron-capture of nuclei with $A<70$. Gaussian
functions were used for $D_{X}(E)$ and $E_{GT}$ was calculated
from Eq.~\rf{13}.}
\end{figure}

In  the evaluation of the allowed electron-capture and
$\beta^+$-decay rates for nuclei  of $A<70$ we have re-adjusted
the parameter $\sigma_N$, imposing again the constraint
$\log ft<6$. The resulting values of $\sigma_N$ and $\eta$  for the two
$\chi^2$-function, with $E_{GT}$ calculated from Eq.~\rf{13}, are
presented in Table \ref{table:2}. Figure~\ref{figure4} shows the
values of $\log(\tau_{1/2}^{calc}/\tau_{1/2}^{exp})$ as a function
of $Q_{\b} A^{1/3}$ for the electron-capture rates
 calculated with the underlined $\sigma_N$ values listed in
Table \ref{table:2}. Similar general features to those
remarked in the $\beta^-$-decay case are obtained.

Also, we briefly discuss the dependence of the $\chi^2$ procedure on the functional
form of the employed strength distribution. Thus, we repeat the
calculations for $\beta^-$-decay and electron-capture rates
using now the lorentzian distribution $D_{X}$, given by the
Eq.~\rf{9}, together with the Eq.~\rf{13} for the GT energy. The
resulting  $\Gamma_N$ energies are shown in
Table~\ref{table:3}, and the corresponding
$log(\tau_{1/2}^{calc}/\tau_{1/2}^{exp})$  values for the
$\beta^-$ emitter nuclei  with $A<70$ exhibit similar $Q_\b
A^{1/3}$ dependence to that shown in Figure~\ref{figure3}.

\begin{figure}[t]
\vspace{-5.cm} \centering
{\includegraphics[width=14cm,height=15cm]{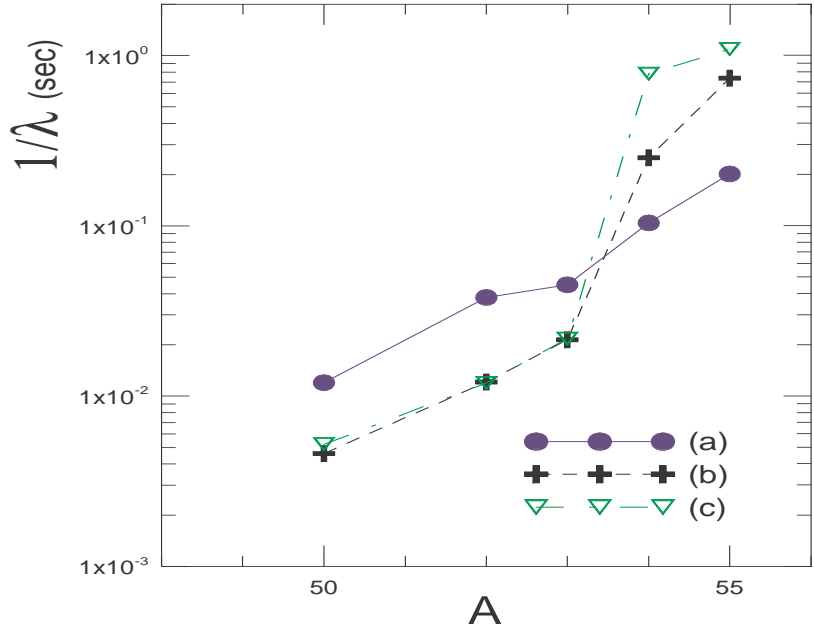}}
\vspace{-.5cm} \caption{\label{figure5}
(Color online) Electron-capture rates for Ni isotopic chain: (a) experimental; (b)
GTBD with gaussian type function; and (c) GTBD with Lorentz type
function. The  energy of the GTR was approximated by Eq.\rf{13}.}
\end{figure}

Figure~\ref{figure5} shows the results for the
electron-capture rates along the  Ni isotopic chain. The
calculations with the gaussian and lorentzian strength functions
turn out to be quite similar to each other and both show
a reasonable agreement with the data.

\subsection{Neutrino-nucleus cross section}

\begin{figure}[t]
\centering
\begin{tabular}{cc}
{\includegraphics[width=7cm,height=7.5cm]{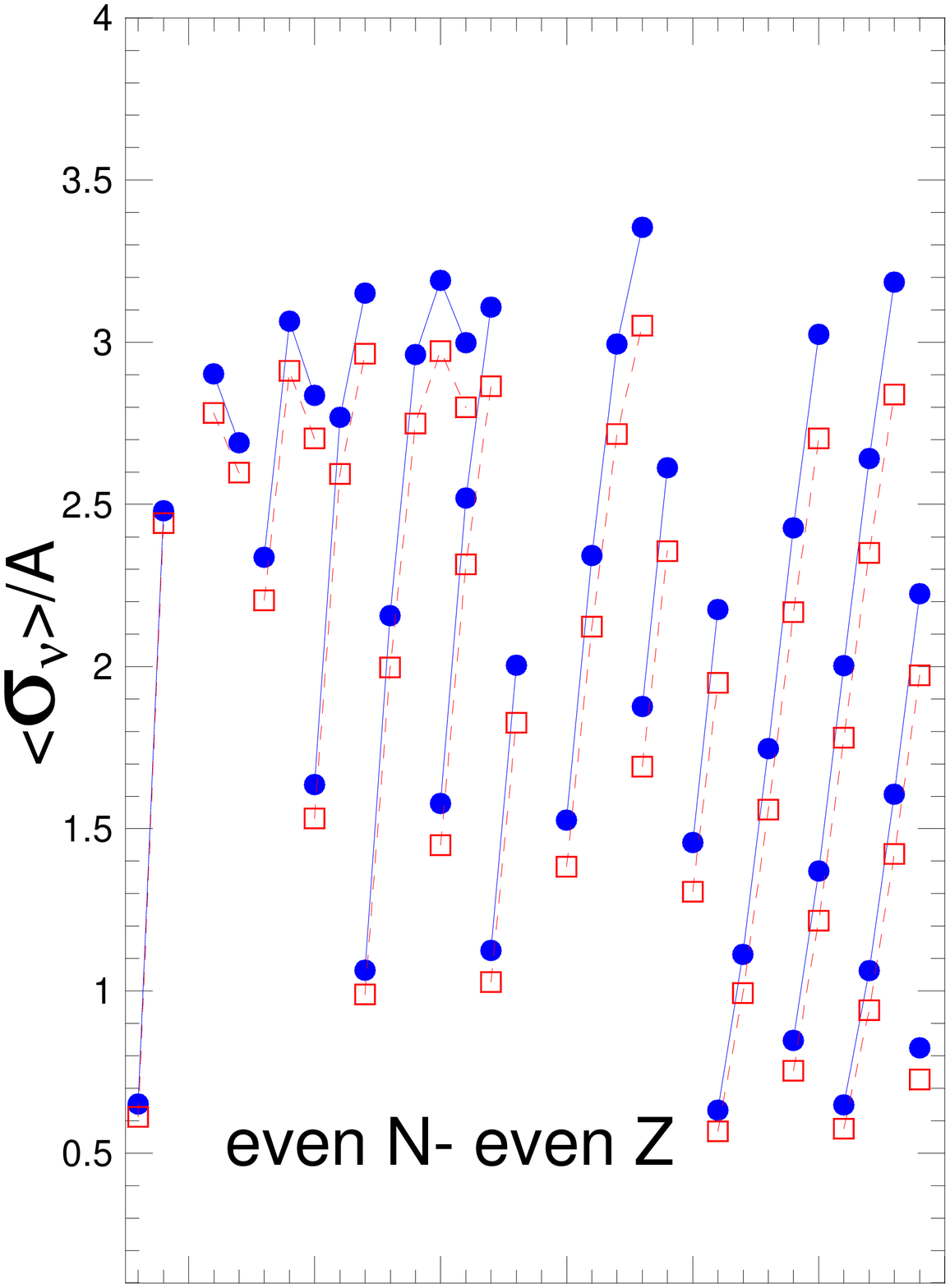}}& \hspace{-.8cm}
{\includegraphics[width=7cm,height=7.5cm]{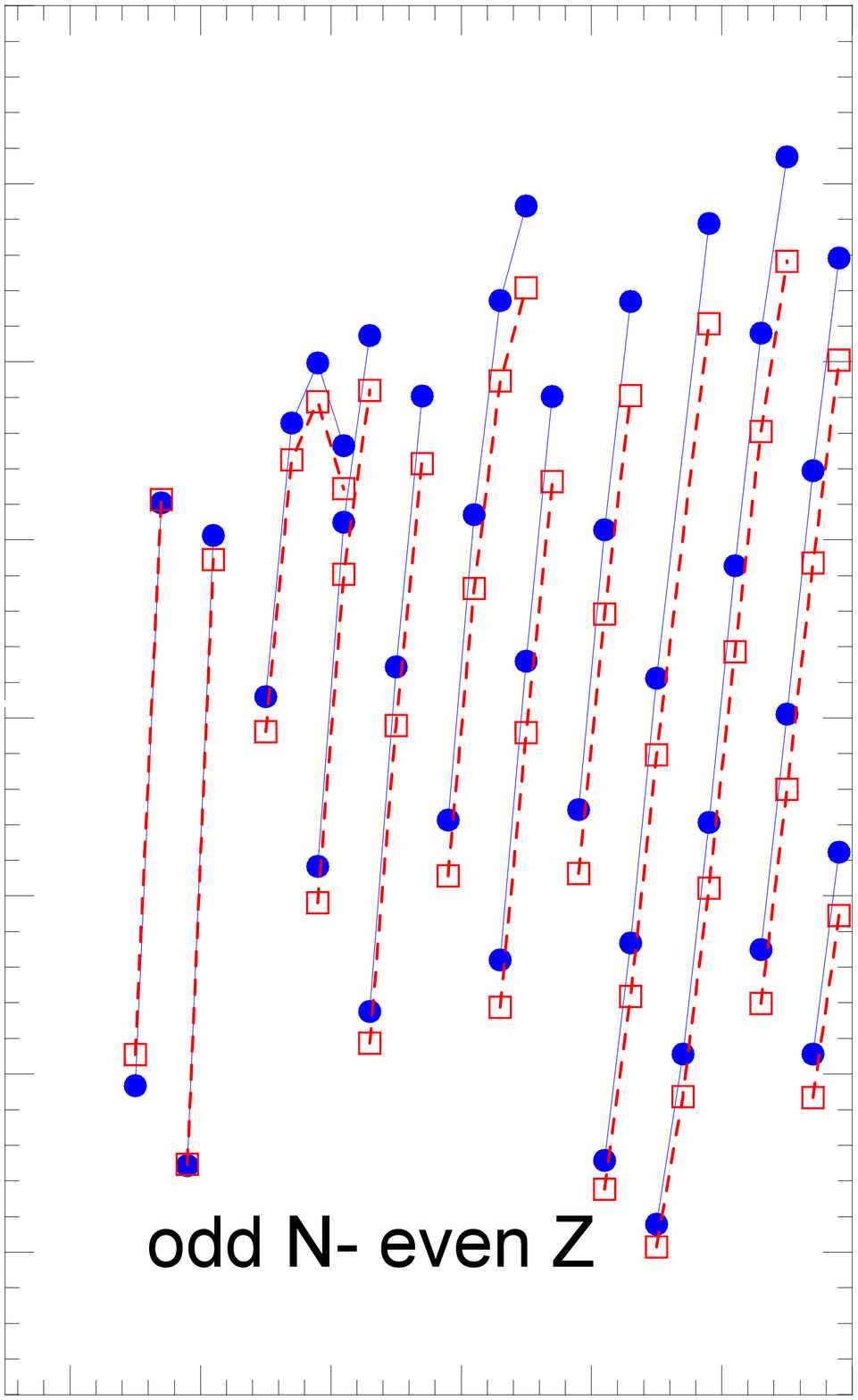}}
\\
{\includegraphics[width=7cm,height=7.5cm]{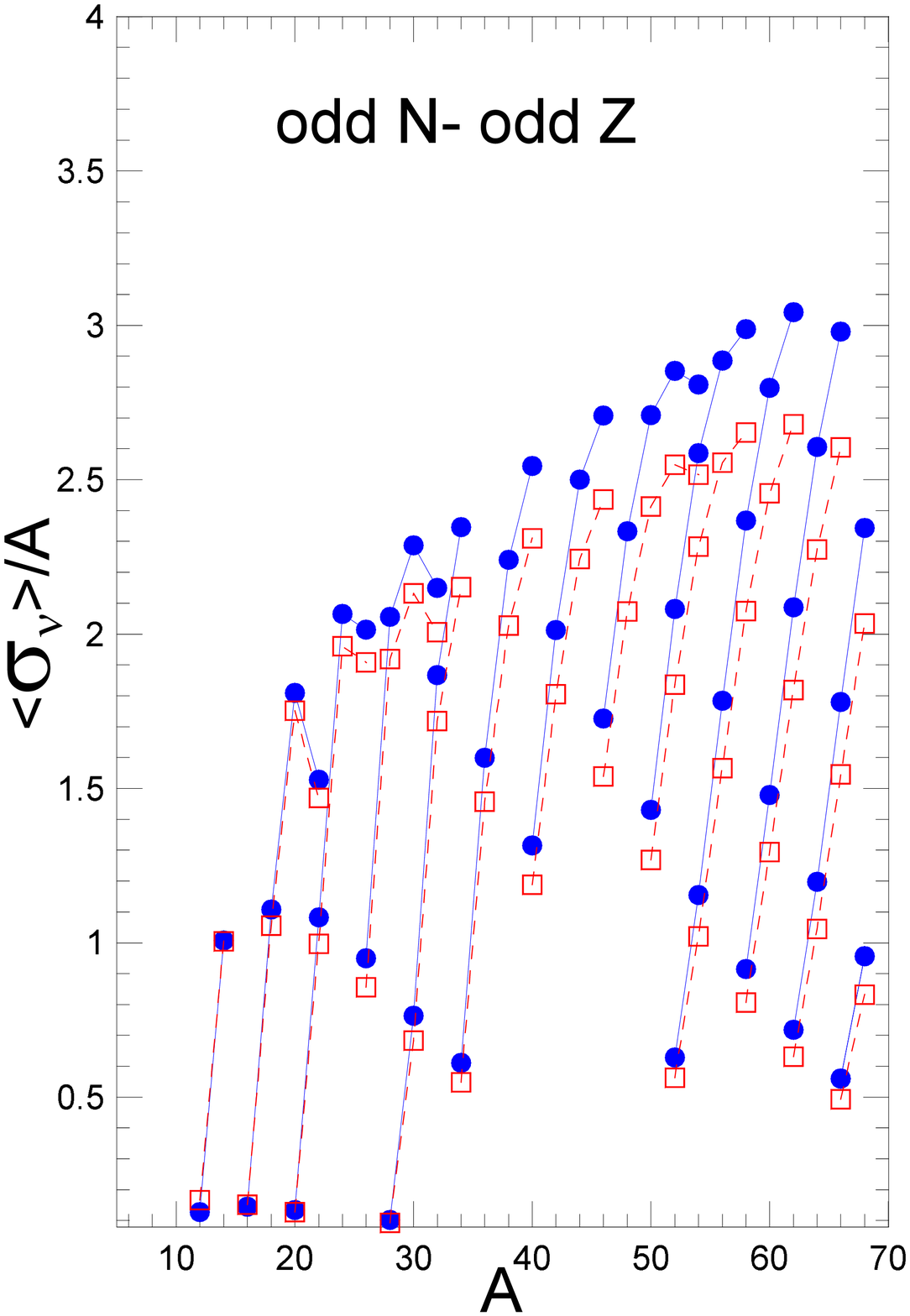}}& \hspace{-.8cm}
{\includegraphics[width=7cm,height=7.5cm]{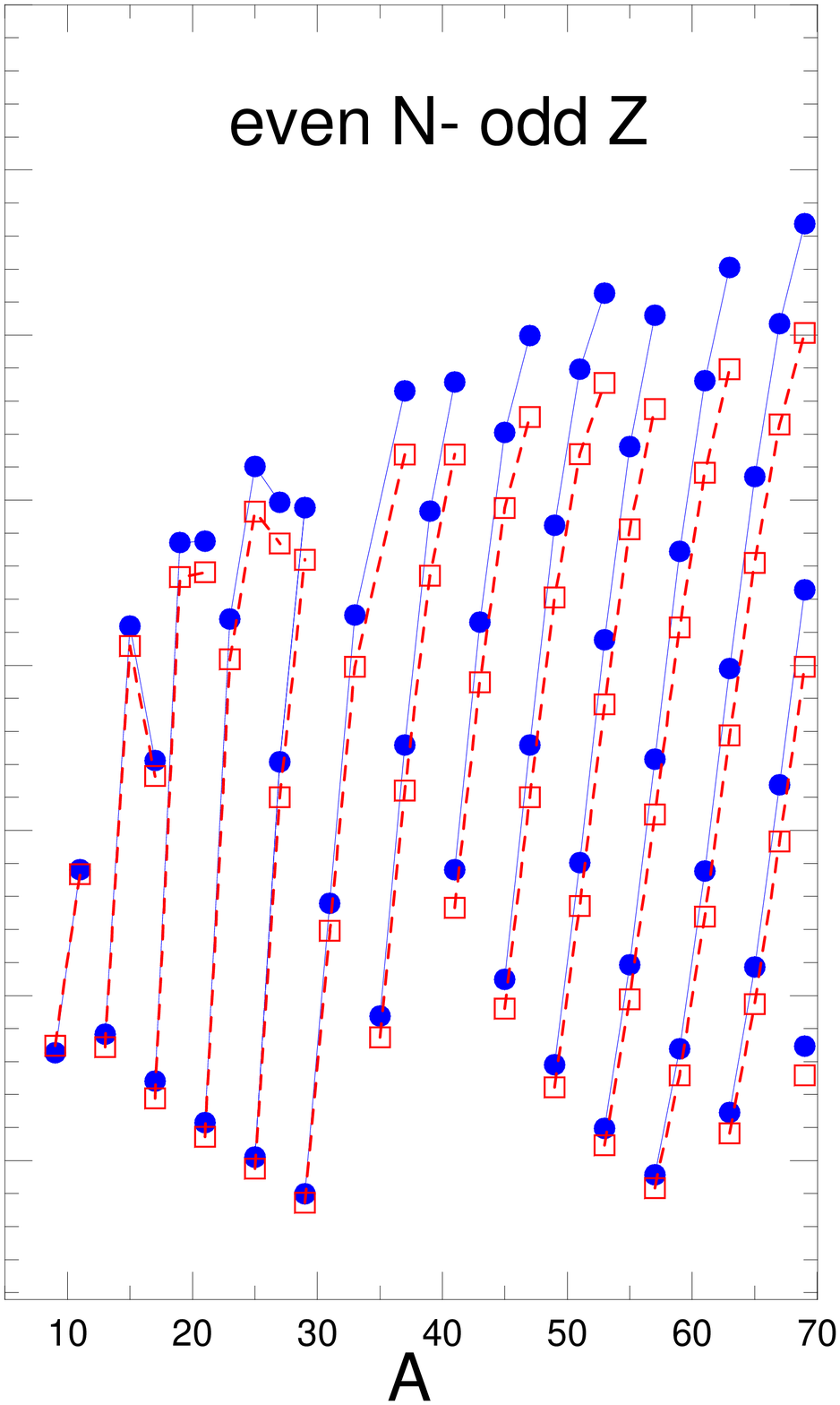}}
\end{tabular}
\vspace{-0.25cm} \caption{\label{figure6} (Color online) Thermal
reduced $\nu_e$-nucleus cross section $\langle
\sigma_\nu \rangle /A$ (in units of $10^{-40}$ cm$^2$) for
$\beta^-$ emitters with $A<70$. Gaussian functions were used for
$D_{X}(E)$. Results obtained  with both approximations for the GTR
are presented; with parameters values given by Eq.~\rf{11}
(filled circles) and by Eq.~\rf{13} (hole squares).}
\end{figure}

\begin{figure}[t]
\vspace{-2.cm}
\centering
\begin{tabular}{cc}
{\includegraphics[width=8cm,height=10cm]{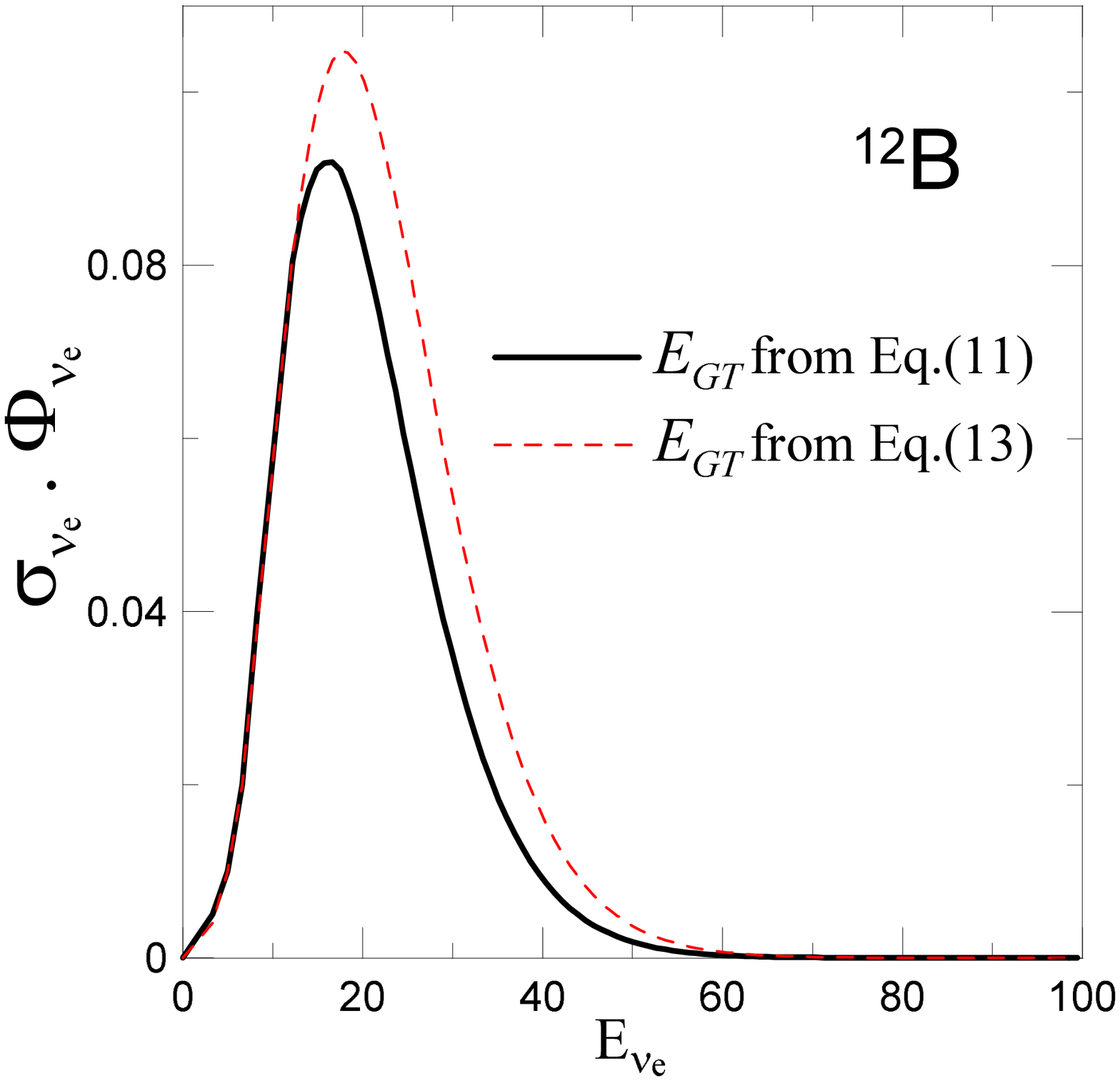}}
&
{\includegraphics[width=8cm,height=10cm]{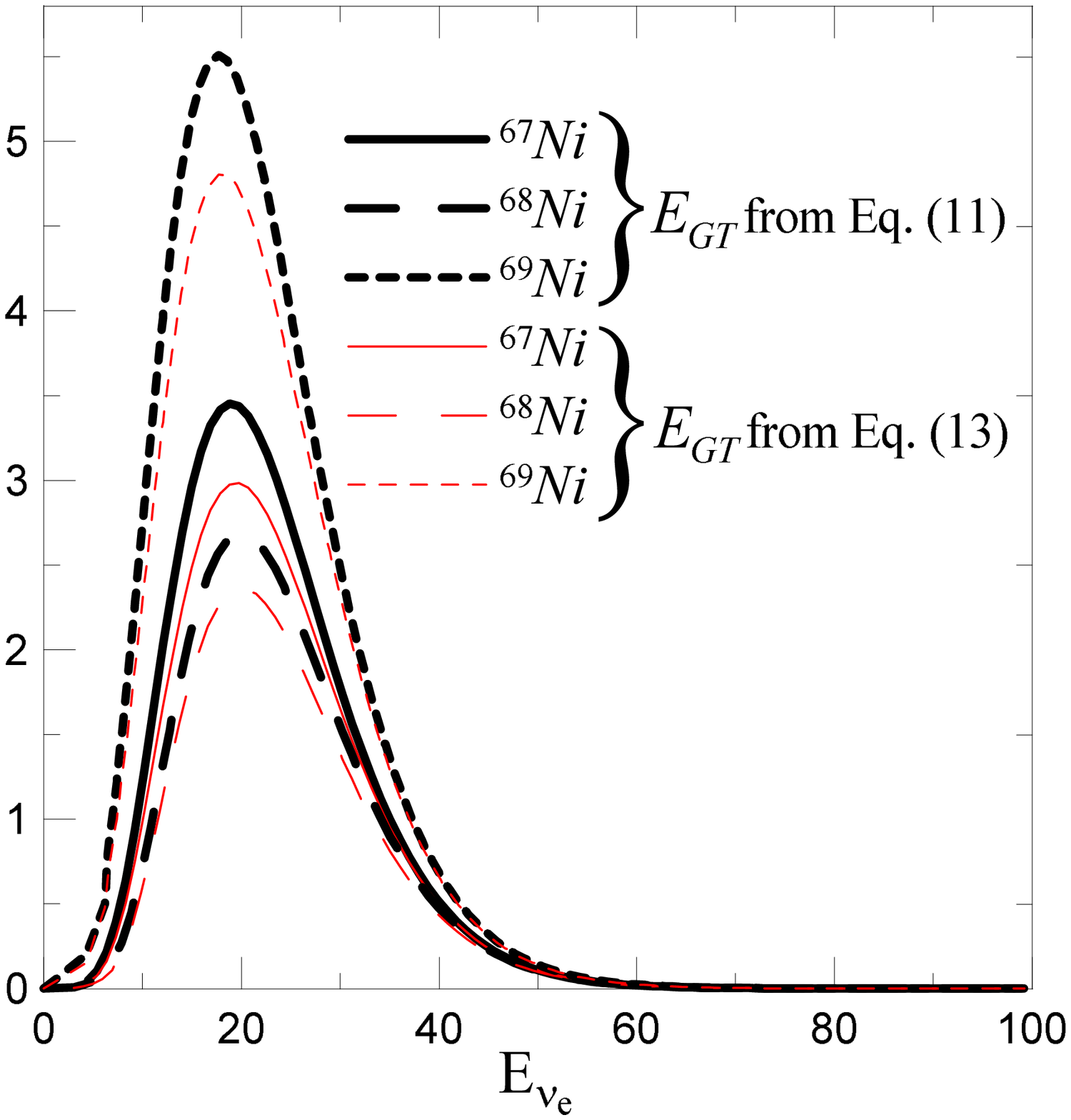}}
\end{tabular}
\caption{\label{figure7} Results for $\sigma(E_\nu) \Phi(E_\nu)$
(in units of $10^{-42}$ cm$^2/$MeV). Gaussian functions were used
for $D_{X}(E)$, and the results with both \rf{11} and \rf{13}
approximations  for $E_{GT}$ are shown; for $^{12}$B (left panel) and
for $^{67, 68, 69}$Ni isotopes (right panel).}
\end{figure}

The  reduced thermal cross section $\langle \sigma_\nu\rangle /A$
of  the four $\beta^-$ emitter families
 was evaluated for the  $A<70$ nuclei with  two  sets of parameters,
$\sigma_N^{*}$ and  $\sigma_N$.
The results, confronted in
Figure~\ref{figure6}, indicate that the Eq.~\rf{13} always yields
smaller values for this quantity than those obtained with the  Eq.~\rf{11},
the difference being more pronounced for $A> 30$.
However, for some isolated light nuclei, the use of
a more realistic GTR  energy increases the cross
section. This is the case of $^{12}$B, for which the product
$\sigma(E_\nu) \Phi(E_\nu)$ is shown in the left
panel of Figure~\ref{figure7}.
The increase of $\sigma(E_\nu)$ arises from the contribution of
the $1^+$ states with energies below the GTR (see Refs.~\cite{Krm05,Sam06}).
As another example, in the right panel of Figure~\ref{figure7} are
shown the results for the Ni isotopes ($^{67}$Ni, $^{68}$Ni and
$^{69}$Ni). One notes that for the three nuclei, the product
$\sigma(E_\nu)\Phi(E_\nu)$ decreases when the energy of the GTR is
moved up. Also, because of the pairing effect, the
cross-section  in $^{68}$Ni presents the lowest value for  both GT
energies.

\begin{figure}[t]
\centering
\includegraphics[width=10cm,height=10cm]{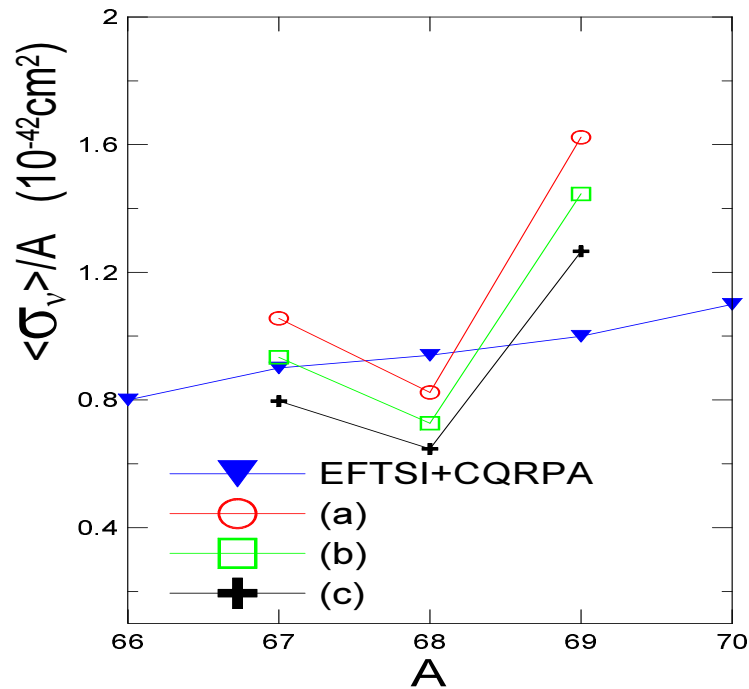}
\vspace{-.5cm} \caption{\label{figure8} (Color online) Comparison
between microscopic ETFSI+CQRPA calculation from Ref.
\cite{Bor00} and our GTNC results for the electronic thermal
reduced neutrino cross section (in units of $10^{-42}$ cm$^2$) for
some Ni isotopes. The results obtained with gaussian strength
functions are shown in (a) with $E_{GT}$ from \rf{11}, and in (b)
with $E_{GT}$ from  \rf{13}. The calculations with Lorentz
distribution and  $E_{GT}$ from  \rf{13} are shown in (c).}
\end{figure}

On the other hand, from Figure~\ref{figure8} it can be seen that our
results for the reduced thermal cross-section in Ni nuclei
emphasizes the odd-even effect when compared with the microscopic
ETFSI+CQRPA calculation~\cite{Bor00}, where  this effect seems to be washed out.
This leads a different
trend of the $\nu_e$-nucleus cross section with respect to $A$.

\begin{figure}[t]
\centering
\begin{tabular}{cc}
{\includegraphics[width=7cm,height=7.5cm]{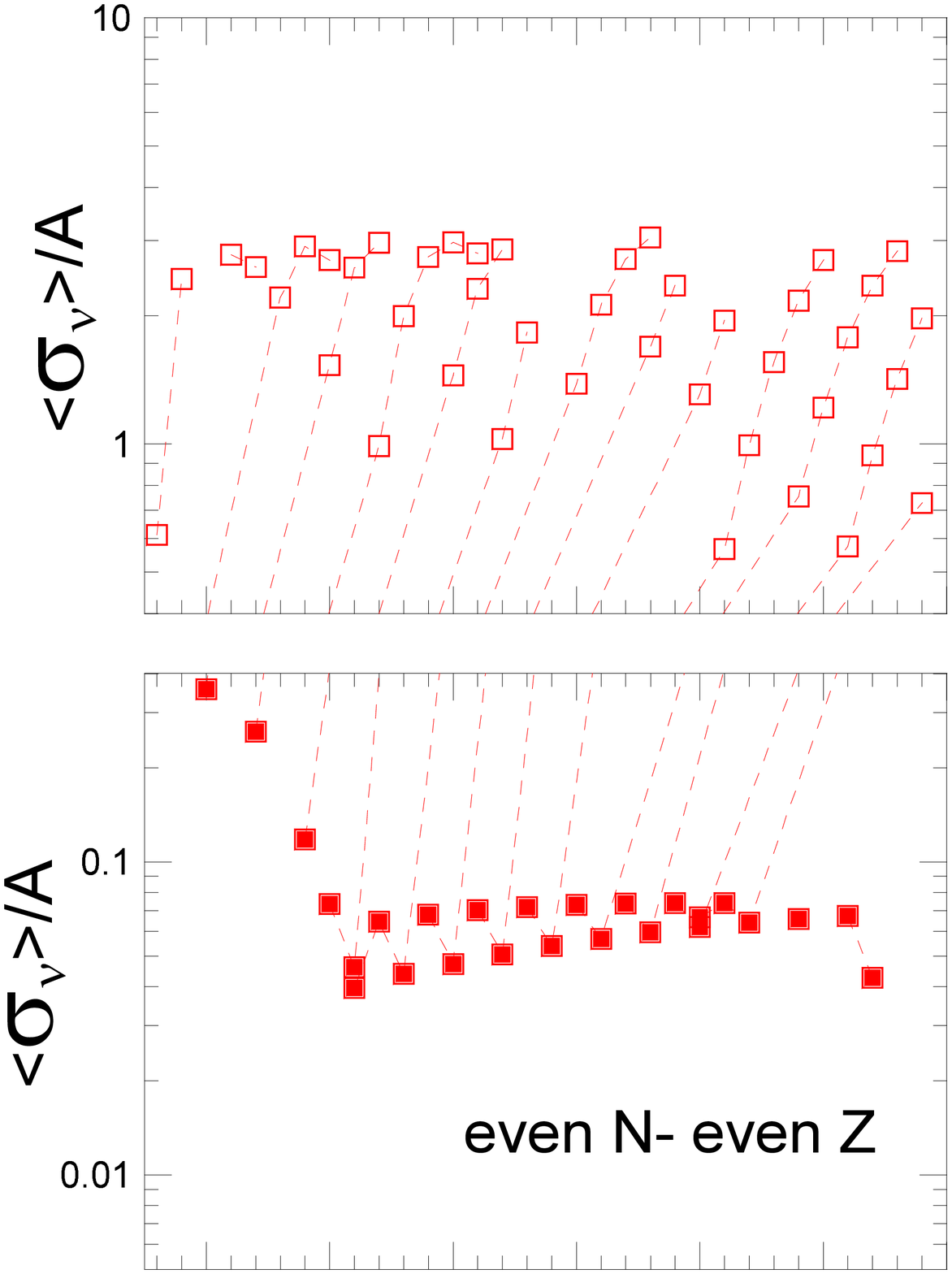}}&
{\includegraphics[width=7cm,height=7.5cm]{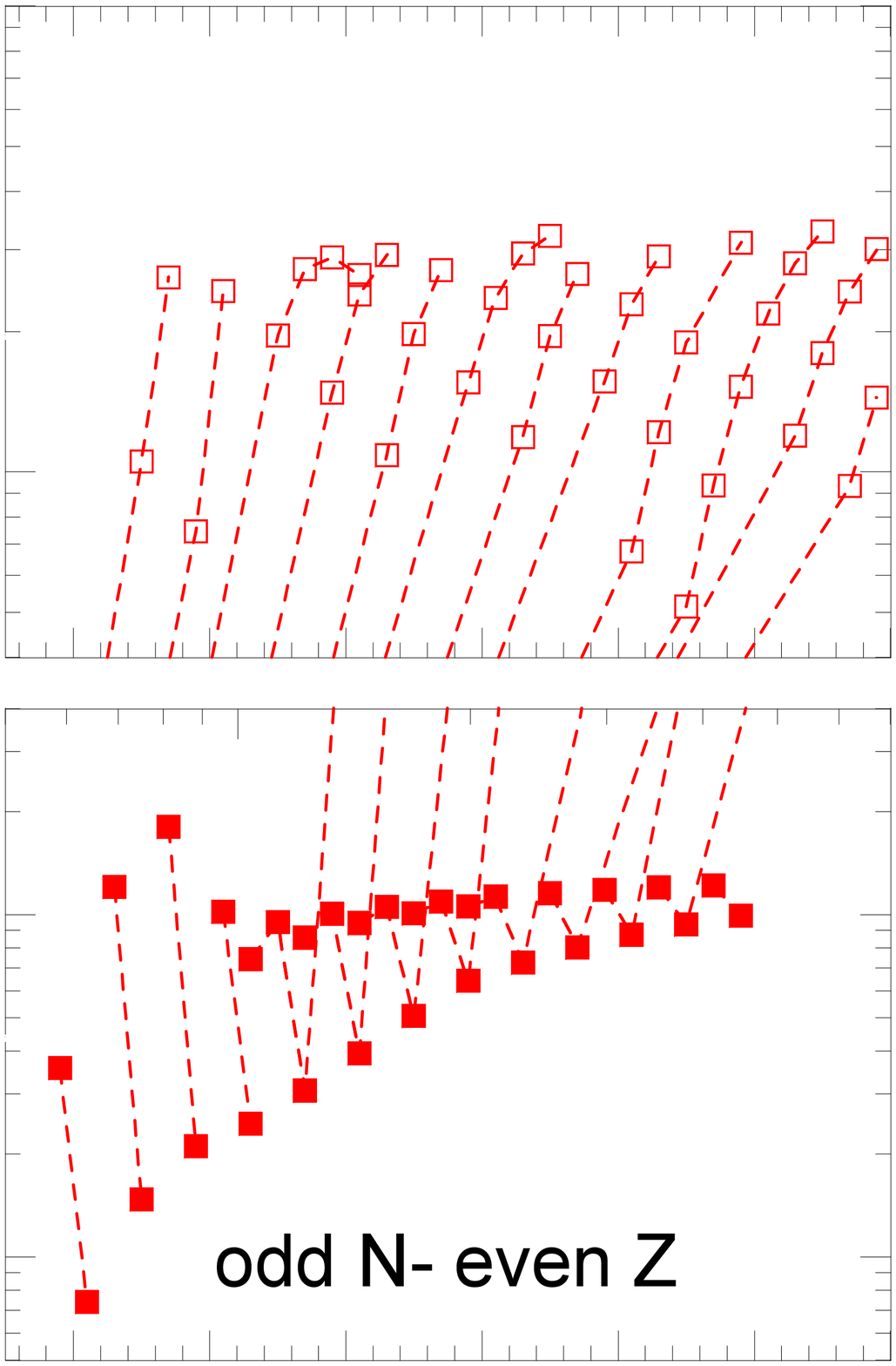}}
\\
{\includegraphics[width=7cm,height=7.5cm]{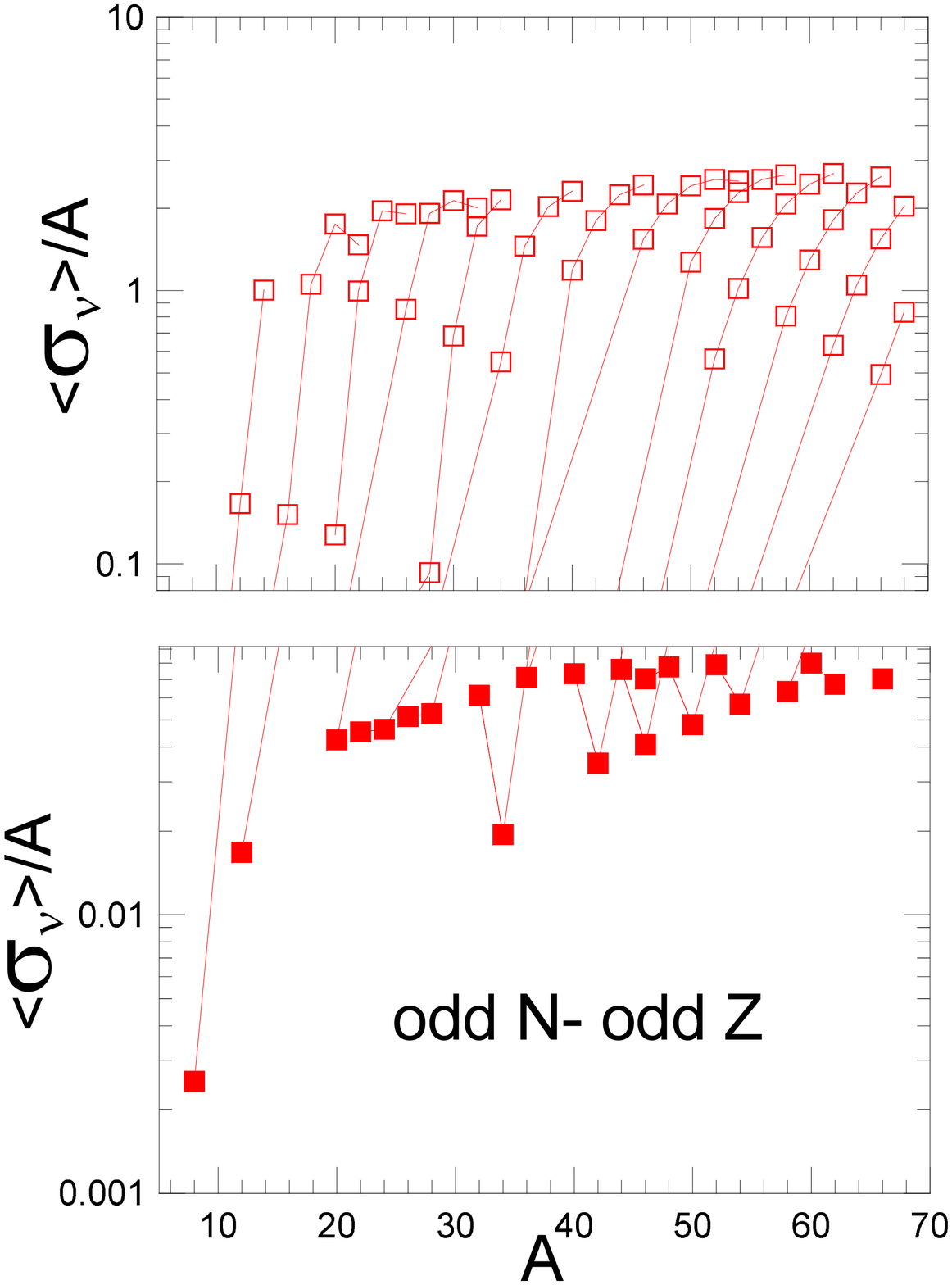}}&
{\includegraphics[width=7cm,height=7.5cm]{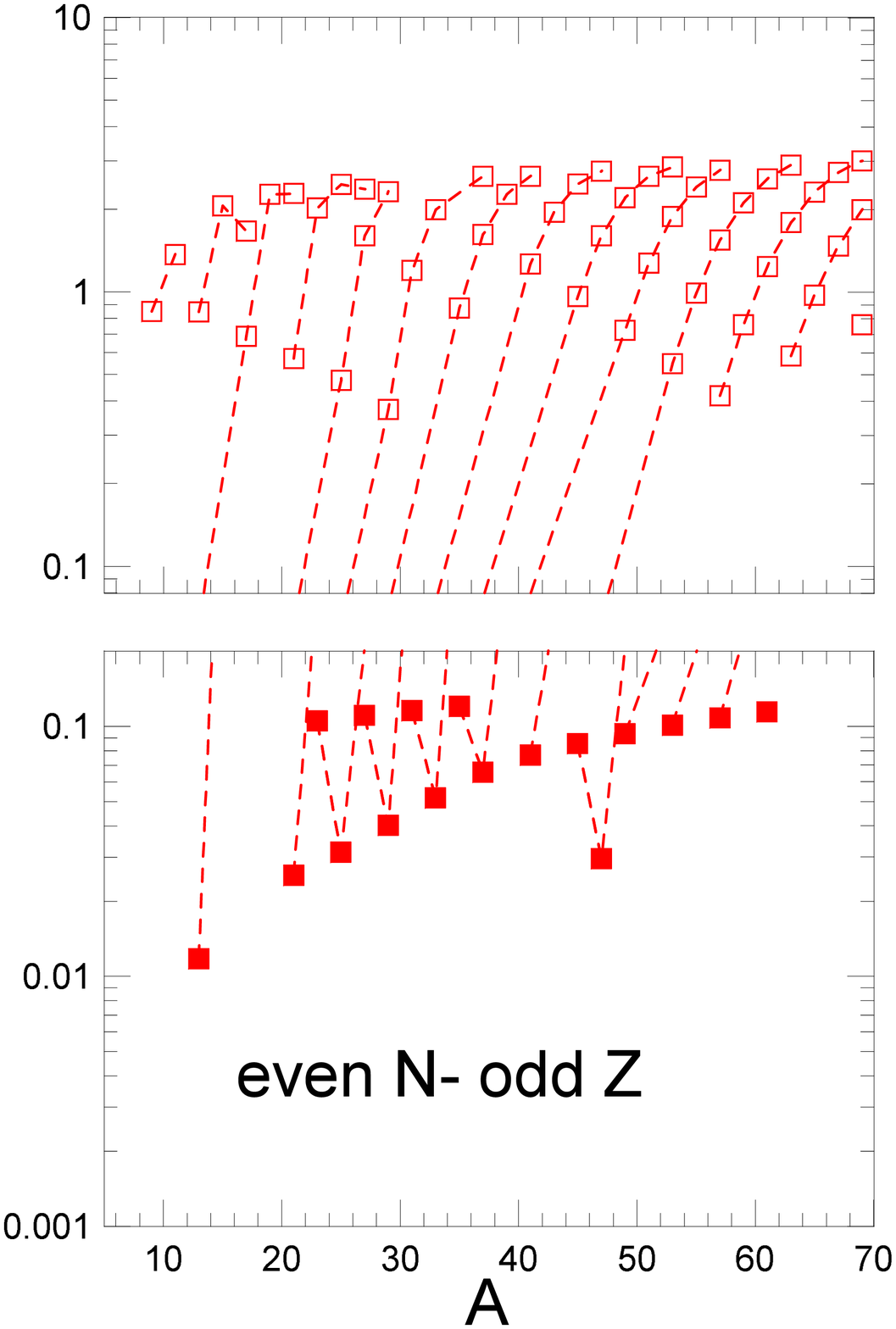}}
\end{tabular}
\vspace{-0.25cm} \caption{\label{figure9} (Color online) Thermal
reduced $\nu_e$-nucleus cross section (in units $10^{-40}$ cm$^2$) for
the  $A<70$ region with the neutrino flux at $T_\nu=4$ MeV. The
Eq.~\rf{13} for $E_{GT}$ was used together with the gaussian
strength function. We present the results for electron-capture
(filled squares) and $\beta^-$ emitters (hole squares).}
\end{figure}

For completeness,  in Figure~\ref{figure9} we present the
results for $\langle \sigma_{\nu} \rangle / A$ obtained with
the GTNC, both for the $\beta^-$decaying nuclei  (with $\sigma_N$
from Table \ref{table:1}), and for the nuclei where take place electron-capture
(with $\sigma_N$ from Table~\ref{table:2}).

It is worth noticing that the  gaussian and lorenzian strength functions
given, respectively, by Eqs. \rf{8} and \rf{9} yield  almost the same results
for the  reduced thermal cross-sections.

At this point it is important to clarify the meaning of the thermal
neutrino flux presented in Eq.~\rf{5}, which we have used for
the calculation of the thermal neutrino-nucleus  cross section
 $\langle \sigma_\nu \rangle$. This neutrino energy flux is given by
a Fermi distribution, \ie Eq.~\rf{5} depending
explicitly on the temperature parameter $T_\nu$. In order
to compare our results with those of Borzov and Goriely \cite{Bor00}
we have used here a  constant temperature $T_\nu=4$ MeV.
However, this situation could not be a realistic one for the supernova
neutrino wind. Neutrinos (and antineutrinos) with
different energies and flavors decouple at different points of the
supernova core and the neutrino spectrum, in fact could be non thermal.
This is due to the non-thermalization of neutrinos through their transport
along hydrodynamics medium evolution~\cite{Kei03,Jac06}. Thus, it could be
interesting to determine the consequences of employing a different
neutrino flux such as a power law flux of the form
\be
\Phi_\nu(E_\nu)={\cal N}_{PL}
\left(\frac{\epsilon_\nu}{\langle \epsilon_\nu
\rangle}\right)^\alpha e^{-\frac{(\alpha+1)\epsilon_\nu}{\langle
\epsilon_\nu \rangle}}.
\label{18}\ee
The parameters $\langle\epsilon_\nu \rangle$ and $\alpha$ are
not fully determined and
here we take $\langle \epsilon_\nu \rangle \approx
3.1514~T_\nu = 12.6056$ MeV, and $\alpha \approx 2.3014$, which
reproduces better the Fermi-Dirac neutrino distribution function
in Eq.~\rf{5} using $T_\nu=4~$ MeV. These parameter values
were obtained in Ref.~\cite{Kei03}.
The normalization constant
${\cal N}_{PL}$ ensures unitary flux between $0$ and $102$ MeV. We
have found that, for all practical purposes, the flux \rf{18}
yields the same results as the thermal flux \rf{5}. This is an
expected result, since  these two fluxes tend to behave
differently only in the tail zone, far away of the integration
interval used to obtain the $\sigma_\nu(E)$ for astrophysical
applications. Some possible deviations in the tail of these fluxes
are important for the rate of nuclear reactions in studies of
astrophysics plasmas \cite{Lis05}.

\section{Summarizing conclusions}

We have briefly revived the original version of the gross theory
for the $\beta$-decay. The main improvement introduced is a
more realistic estimate for the location of the GTR energy
peak, $E_{GT}$. After fixing the free parameter
of our model ($\sigma_N$ or $\Gamma_N$, depending on the
parametrization adopted for the strength function) we have
calculated the $\beta^-$-decay  and electron-capture rates. A
careful selection of input data for $A < 70$ nuclei, with  small
error bars in the measured  half-lives, has been done in order to fix
the model parameters in the fitting procedure.
The model can be extended to
the $A > 70$  nuclei, as well as to the transuranic nuclei, which
are of interest for the study of the $r$-process in supernova.
The first- and second-forbidden weak
processes could play an important role in the exotic nuclei within
this nuclear mass region. But, these transitions  can be easily
included in the  gross theory framework, as has been done already
by Nakata {\it et al.} \cite{Nak97} within the semi-gross theory
\footnote{A relation analogous to \rf{13} was also derived for the
first forbidden charge-exchange resonances~\cite{Krm83}, which
is quite different to the one used in Ref.~\cite{Nak97}. Thus  it
might be more appropriate  to employ \cite[Eqs. (3.11) and
(3.12)]{Krm83}, instead of \cite[Eq. (48)]{Nak97}.}.

The present results are encouraging, in the sense that the gross
theory could be able to describe in a systematic way, not only the
nuclear  properties along the $\beta$-stability line, but also for
exotic nuclei involved in presupernova composition.
In particular, the results for the reduced thermal cross section
 $\langle \sigma_\nu\rangle /A$
in the region $A < 70$ are in fair agreement with previous
calculation performed within more refined microscopic
models, \ie the ETFSI+CQRPA model \cite{Bor00}. The difference
between the two descriptions could be attributed to the use of
the Fermi gas model which contains more degrees of
freedom that the EFTSI+CQRPA. Consequently, in general,
$\sigma_\nu(E_\nu)$ calculated with the Fermi gas model
leads to values higher than those obtained with microscopical nuclear
models~\cite{Mon69,Kol94,Ath98},
particularly for light or intermediated nuclei
(see, for instance, the results
for the $\nu-{^{12}}$C reaction  shown in  \cite[Fig. 2)]{Nak97}
and \cite[Fig. 32]{Ath98}).

The important aspect of the recent $r$-process calculations is
that they take into account the neutrino-rich environment
in supernova explosion, where the  $\nu_e$-nucleus reaction are in
competition with the $\beta$-decay processes~\cite{Ter04}. To
address this type of calculation we have evaluated the
cross-section  $\sigma_\nu(E_\nu)$ within the GTNC model, folded
with a temperature dependent neutrino flux.

Finally, we want to remark once more the simplicity of the present
model, which we are planing to extended in the near future to the
$r$-process nuclei region, as well as to evaluate the
isotopic abundances in presupernova scenario.

\section*{Acknowledgments}

A.R.S. thanks the financial support of FAPERJ (Rio de Janeiro,
Brasil) and from Texas A\&M University-Commerce. Two of us (C.A.B.
and F.K.) are members of CONICET (Argentina). S.B.D thanks the
partial support from CNQq, Brazil. F.K. acknowledges  the support
of  FAPESP (S\~ao Paulo,  Brazil). The authors thank C. Bertulani
for careful reading and revision of the manuscript.

\newpage

\begin {table}[t]
\begin{center}
\caption {Standard deviations $\sigma_N$ (in units of MeV) and
 mismatch factors  $\eta$ for $\beta^-$-decay. Gaussian
single-particle strength probability function $D_{X}(E)$ was
adopted. $\sigma_N$ and $\eta$ ($\sigma_N^{*}$ and $\eta^*$)
indicate the results obtained with $E_{GT}$ approximated from Eq.
\rf{13} (Eq. \rf{11}). Parenthetically are shown the values
obtained by Takahashi {\it et al} \cite{Tak69} for a different data
set of nuclei. The electronic neutrino cross-section are evaluated
with the  underlined values of $\sigma_N$.}
\label{table:1}
\newcommand{\cc}[1]{\multicolumn{1}{c}{#1}}
\renewcommand{\tabcolsep}{.7pc} 
\renewcommand{\arraystretch}{1.2} 
\bigskip
\begin{tabular}{  c|c|cccc|cccc}\hline
$N-Z$      &&\multicolumn{4}{c|}{$\chi_A^2$}
            &\multicolumn{4}{c}{$\chi_B^2$}\\
(parent)   &$N_0$
              &$\sigma_N^{*}$&$\eta^{*}$
                   &$\sigma_N$&$\eta$
                        &$\sigma_N^{*}$&$\eta^{*}$
                              &$\sigma_N$&$\eta$
                                   \\\hline\hline
odd-odd    &54    &13.3~(5.0)&9.7~(45.5)&17.6&10.7  & 8.6&10.6&\underline{15.8}&10.7 \\
even-even  &43    &13.5~(4.5)&9.3~(12.9)&16.3&10.0  & 9.7&14.6&\underline{15.8}&10.0 \\
odd-even   &40    &13.0~(5.1)&6.1~ (9.4)&16.8& 6.4  & 4.1&15.6&\underline{ 7.2}& 9.8 \\
even-odd   &55    &13.8~(5.1)&7.3~ (6.5)&17.6& 7.7  &10.4& 7.4&\underline{16.5}& 7.7 \\
\hline\hline
\end{tabular}\end{center}\end {table}

\newpage

\begin {table}[h]
\begin{center}
\caption { Standard deviations $\sigma_N$ (in units of MeV) and
 mismatch factors  $\eta$ for  $\beta^+$ decay and electron-capture. Gaussian
single-particle strength function $D_{GT}$ was used. The energy
$E_{GT}$ has been evaluated from Eq. \rf{13}. The remaining
notation is the same as in Table \ref{table:1}. No minimum has
been found for the $\chi_A^2$-function in the case of even-even
parent nuclei.}
\label{table:2}
\newcommand{\cc}[1]{\multicolumn{1}{c}{#1}}
\renewcommand{\tabcolsep}{1.pc}  
\renewcommand{\arraystretch}{1.2}
\bigskip
\begin{tabular}{  c|c|c|c|c|c}\hline
$N-Z$      &  &\multicolumn{2}{c|}{$\chi_A^2$}
                          &\multicolumn{2}{c}{$\chi_B^2$}
\\\hline
(parent)   &$N_0$
               &$\sigma_N$
                               &$\eta$
                                      &$\sigma_N$
                                           &$\eta$
\\\hline\hline
odd-odd    &23       & 9.7 &10.7 &\underline{10.4}&10.7\\
even-even  &24       &  -  &   - &\underline{ 9.9}& 5.2\\
odd-even   &32       &12.5 & 6.4 &\underline{11.8}& 9.8\\
even-odd   &17       &12.2 & 7.7 &\underline{12.2}& 7.7\\
\hline\hline
\end{tabular}\end{center}\end {table}

\newpage

\begin {table}[h]
\begin{center}
\caption {Standard deviations $\sigma_N$ (in units of MeV) and
 mismatch factors  $\eta$ for $\beta^-$-decay and electron-capture, obtained from the minimization
of the $\chi^2_B$-function. Lorentzian single-particle strength
probability function was used. The energy $E_{GT}$ has been
evaluated from Eq. \rf{13}.} \label{table:3}
\newcommand{\cc}[1]{\multicolumn{1}{c}{#1}}
\renewcommand{\tabcolsep}{1.pc} 
\renewcommand{\arraystretch}{1.2} 
\bigskip
\begin{tabular}{c|c|c|c|c|c|c}\hline
$N-Z$      &\multicolumn{3}{c|}{$\beta^-$ decay}
           &\multicolumn{3}{c}{$e^-$-capture}
\\\hline
(parent)   &$N_0$
               &$\Gamma_N/2$
                  &$\eta$
                         &$N_0$
                              &$\Gamma_N/2$
                                   &$\eta$
\\\hline\hline
odd-odd    &54&15.2&12.7&23 & 9.8&11.3 \\
even-even  &43&15.4&11.6&24 & 9.4&6.5  \\
odd-even   &40&8.5 &8.0 &32 &11.3&6.4  \\
even-odd   &55&15.7&8.6 &17 &11.5&8.0  \\
\hline\hline
\end{tabular}\end{center}\end {table}

\newpage

\newpage

\newpage

\newpage

\newpage

\newpage

\newpage

\newpage

\newpage

\end{document}